\documentclass[runningheads]{llncs}

\usepackage[T1]{fontenc}
\usepackage[utf8]{inputenc}
\usepackage{amsmath,amssymb}
\usepackage{booktabs,tabularx,multirow,array}
\usepackage{graphicx}
\usepackage{tikz}
\usetikzlibrary{patterns}
\usepackage{algorithm}
\usepackage{algpseudocode}
\usepackage{xcolor}
\usepackage{microtype}
\usepackage{xurl}
\usepackage[hidelinks]{hyperref}
\usepackage[capitalise,noabbrev]{cleveref}

\newcommand{\sha}{SHA-256}
\newcommand{\Sig}[1]{\Sigma_{#1}}
\newcommand{\sig}[1]{\sigma_{#1}}
\newcommand{\MAJ}{\mathrm{MAJ}}
\newcommand{\IF}{\mathrm{IF}}
\newcommand{\cv}{\mathit{cv}}
\newcommand{\Feven}{F_{\mathrm{even}}}
\newcommand{\Fodd}{F_{\mathrm{odd}}}
\newcommand{\ptail}{p_{\mathrm{tail}}}

\newcommand{\pclaw}{p_{\mathrm{claw}}}
\newcommand{\hex}[1]{\texttt{0x#1}}

\begin{document}

\title{Free State Words, Claws and Lists: A Collision Attack on 39-Step \sha{} in One MiB}
\titlerunning{Free State Words, Claws and Lists}

\author{Dariia Porechna\inst{1}\thanks{Corresponding author.} \and Jeremy Benisek\inst{2} \and Mitchell Travers\inst{3} \and Brian Carpenter\inst{4} \and Pontus Olovsson\inst{5} \and Damian Kuthoore \and Jordan Newman\inst{6}}
\authorrunning{D. Porechna et al.}
\institute{EternaX Labs \email{dariia.p@eternax.ai}
\and Don't Panic Business Technologies \url{https://dontpanic.biz}
\and Soulbis / agentprivacy \email{mitchell@soulbis.com}
\and Deep Fork Cyber \url{https://deepforkcyber.com}
\and Link\"oping University \email{pontus.olovsson@liu.se}
\and Omnispace Technologies \email{Jordan@omnispacetechnologies.com}}

\maketitle

\begin{abstract}
Li \emph{et al.} recently gave the first collision attack on 38-step \sha{} and a 39-step differential characteristic, but judged their method ineffective at 39 steps, where it costs $2^{168}$. We show that this characteristic supports a two-block collision attack on 39-step \sha{} that costs about $2^{98.7}$ compression-equivalents per expected collision with one MiB of memory, where every operation is counted on a bit-sliced machine on which one 39-step compression costs 205.6 operations; for any dependence between the trials' tails, the expected cost of the first collision is at most $2^{98.8}$. The attack rests on three observations. At 39 steps the first two state words carry no conditions, and leaving both free turns the 32-bit condition on $W_{16}$ into a claw of a single 32-bit map, found with about $2^{11}$ evaluations. The map from the chaining value and the free words to the message words that drive the attack is a bijection, which makes the side conditions on $W_8$ and $W_9$ exact. The next two state words reach $W_{16}$ only through one 32-bit function; on a fibre of that function the claw is unchanged, and each claw tests the condition on $W_{17}$ against 9.7 million second blocks. The construction does not depend on the characteristic, which sets only the price of the free words; measured on the eleven published characteristics of 36 to 39 steps of SHA-256 and SHA-512, the claw pays when the local collision starts at $W_7$ or later, and not earlier. The cost rests on stated heuristic premises, chiefly the measured probability, about $2^{-74.5}$, that steps 18 to 38 end in a collision; the attack has not been run at full scale, and no collision has been found. Timed on one machine, the attack is 4.5 to 5.5 bits cheaper than the joint message-word pinning of Gong \emph{et al.}, which needs a 17\,GiB table, on either tail event. On 39-step SHA-512, the same list, searched bit by bit against the conditions on $W_{17}$, lowers the best published estimate from $2^{131.3}$ to about $2^{101}$ compressions per expected collision.

\end{abstract}

\section{Introduction}\label{sec:intro}

The SHA-2 family has withstood the techniques that broke MD5 and SHA-1~\cite{WFLY04,SBK+17}, and its full versions retain a substantial security margin against collision attacks. Progress on step-reduced \sha{} has come in small increments. By 2008, collision attacks had reached 24 steps~\cite{SS08,IMPR08}. Automated characteristic search~\cite{MNS11} and new local-collision constructions~\cite{MNS13} brought the collision frontier to 27 and then 31 steps, where it stayed for more than a decade; a practical 31-step collision followed in 2024~\cite{LLW+24b}. In 2026, SAT-based characteristic search and memory-efficient frameworks for satisfying conditions advanced it rapidly: theoretical collision attacks on 36 and 37 steps~\cite{ZMLM26}, a practical collision attack on 35 steps~\cite{LLWS26}, and, by Li \emph{et al.}~\cite{LZL+26}, the first collision attack on 38-step \sha{} at $2^{104.3}$, together with the first practical 36-step colliding pair and the first effective attack on 39-step SHA-512, at $2^{178}$.

For 39-step \sha{}, Li \emph{et al.} find a differential characteristic and a semi-free-start colliding pair, but conclude that their attack is ineffective. In their framework, a dense solution of the early steps is fixed once, each new chaining value is connected to it through the first message words, and every condition from step 16 onward is satisfied by trial. At 39 steps there are 168 such conditions, so the attack would cost $2^{168}$, ``worse than the generic birthday attack''~\cite[Sect.~4.3]{LZL+26}.

\paragraph{Setting.} The attacks in this paper were developed as filings on HashSmash, a public board launched by Eigen Labs and Shielded Labs, on which participants file reproducible attack packages against step-reduced hash functions~\cite{HashSmashSite,HashSmash}. An automated judge checks each package, runs its declared experiments in a sandbox and scores it in a fixed word-RAM cost model (\cref{sec:costmodels}); a human reviewer then decides on acceptance. The judge tests the attack's components without running it at full scale. Filings by four participants between 16 and 19 October 2026 built the attack with two free state words (\cref{sec:two}), and filings between 27 October and 18 November 2026 improved it and, from 8 November, extended it to four (\cref{sec:list,sec:attack}). The attack of this paper is the one implemented in filing 466760ac~\cite{PR1809}, which we call the \emph{reference implementation}; that of the attack with two free words is~\cite{PR574}. \Cref{app:provenance} records the contribution of each filing. The measurements on 39-step SHA-512 (\cref{sec:sha512}) are not a filing. Costs are stated in compression-equivalents, a unit in which every operation of the attack is charged at the price of a compression on the same machine (\cref{sec:costmodels}), and in the board's model.

\paragraph{Concurrent work.} Gong, Wei, Zhang and Wei~\cite{GWZW26}, received by the Cryptology ePrint Archive on 11 November 2026, attack 39-step \sha{} on the same characteristic by \emph{joint message-word pinning}, which solves the same two equations by look-ups in a 17\,GiB table, and report $2^{108.8}$ timed scalar compressions per expected collision. \Cref{sec:pinning} compares the two attacks.

\paragraph{Our contributions.}
\begin{enumerate}
\item \emph{A collision attack on 39-step \sha{} at $2^{98.7}$ in one MiB.} With the claw search implemented as a bit-sliced program in which every operation is counted, and steps 18 to 38 ending in a collision with a measured probability of about $2^{-74.5}$, the attack costs about $2^{98.7}$ compression-equivalents per expected collision with one MiB of memory (\cref{sec:attack}, \cref{tab:summary}). For any dependence between the tails of its trials, the expected cost of the first collision is at most $2^{98.8}$ (\cref{prop:success}). Timed on one machine, it is 4.5 to 4.6 bits cheaper than joint message-word pinning~\cite{GWZW26} on the strict tail event on which joint pinning is priced, and 5.3 to 5.5 bits cheaper on the event on which the list attack is priced, with one MiB of memory against 17\,GiB (\cref{sec:pinning}).
\item \emph{Freeing state words turns rows 16 and 17 into a claw and a list.} At 39 steps the local collision starts at $W_8$, so the first two state words $A_0$, $A_1$ carry no differential condition. With both free, the condition on the expanded word $W_{16}$ separates as $F(a_0)=F(u)$ for a single 32-bit map $F$ (\cref{prop:claw}), and a distinguished-point search finds claws of $F$ at about $2^{11}$ evaluations each, instead of $2^{32}$ trials. The transplant onto a random first block is a bijection on 320 bits (\cref{thm:bijection}), so the side conditions on $W_8$ and $W_9$ hold with probability exactly $585/2^{19}$ (\cref{lem:side}). The next two state words, $A_2$ and $A_3$, reach $W_{16}$ only through a 32-bit function $h$. On a fibre of $h$ the claw is unchanged, and row 17 becomes an equation in the same map, which each claw tests against a list of 9{,}701{,}592 second blocks (\cref{sec:list}).
\item \emph{The construction works for any characteristic, and its price decides where it pays.} \Cref{prop:claw,lem:four} depend neither on the characteristic nor on the word size; the characteristic sets only the price of the free state words. Measured on all eleven published characteristics of 36 to 39 steps of SHA-256 and SHA-512, that price makes the claw pay when the local collision starts at $W_7$ or later, and exceed the whole of row 16 when it starts earlier (\cref{sec:criterion}). On 39-step SHA-512, where row 17 admits $2^{30.4}$ values of $W_{17}$, a depth-first search over the bits of $a_2$ finds the entries of the list that meet row 17 without visiting the others. It gains 29 bits over single-word pinning~\cite{GWZW26} timed with the same code, which lowers the estimated cost from $2^{131.3}$ to about $2^{101}$ compressions per expected collision (\cref{sec:sha512}).
\item \emph{The premises are stated, and each one that can be measured is.} The cost rests on declared premises (\cref{sec:verification}). The estimator of the uncontrolled steps is unbiased (\cref{prop:smc}), and the bound on the first collision holds for any joint law of the tails (\cref{prop:success}). Measured in pre-registered runs on real claws of the walk, the tail is at most $0.025$ bits below its value on uniform draws at 95\% confidence. The event on which the attack is priced is about 6 bits more likely than the strict event of the characteristic. About 3.7 bits of the difference come from collisions that break sign conditions of $\Sig{1}$ and $\IF$ at steps 20 and 21 while keeping the characteristic's state differences, and the rest from collisions that leave the characteristic after row 23 (\cref{sec:gap}).
\end{enumerate}

\begin{table}[t]
\centering
\caption{Selected collision attacks on step-reduced \sha{}, adapted from~\cite[Table~1]{LZL+26}. Times are per expected collision. They are in different units: estimates in compression calls for~\cite{LZL+26} and the earlier attacks, timed scalar compressions on the authors' machine for~\cite{GWZW26}, and compression-equivalents (\cref{sec:costmodels}) for this work. At 39 steps, steps 18 to 38 are priced on different events: strict, the strict event of the characteristic; $K_P$, the carrier (\cref{sec:carrier}), about 6 bits more likely. On one event and timed on one machine, the list attack is 4.5 to 5.5 bits cheaper than joint pinning (\cref{tab:timing}). SFS: semi-free-start.}\label{tab:summary}
\setlength{\tabcolsep}{5pt}
\begin{tabular}{@{}lllll@{}}
\toprule
Type & Steps & Time & Memory & Reference \\
\midrule
\multirow{14}{*}{collision}
& 24 & practical & & \cite{SS08,IMPR08} \\
& 27 & practical & & \cite{MNS11} \\
& 28 & practical & & \cite{MNS13} \\
& 31 & $2^{65.5}$ & $2^{34}$ & \cite{MNS13} \\
& 31 & $2^{49.8}$ & $2^{48}$ & \cite{LLW24a} \\
& 31 & practical & & \cite{LLW+24b} \\
& 35 & practical & & \cite{LLWS26} \\
& 36 & practical & & \cite{LZL+26} \\
& 37 & $2^{79.1}$ & negligible & \cite{LZL+26} \\
& 38 & $2^{104.3}$ & negligible & \cite{LZL+26} \\
& 38 & $2^{90.9}$ & 17\,GiB & \cite{GWZW26} \\
& 39 & $2^{168}$ (ineffective) & & \cite{LZL+26} \\
& 39 & $2^{108.8}$ (strict) & 17\,GiB & \cite{GWZW26} \\
& 39 & $2^{98.7}$ ($K_P$) & $2^{20}$ bytes & this work \\
\midrule
\multirow{2}{*}{SFS collision}
& 38 & practical & & \cite{MNS13} \\
& 39 & practical & & \cite{LLW24a} \\
\bottomrule
\end{tabular}
\end{table}

\paragraph{Organization.} \Cref{sec:prelim} fixes notation and recalls \sha{}, the two-block framework and parallel collision search. \Cref{sec:transplant} describes the 39-step characteristic, the free state words and the transplant. \Cref{sec:claw} shows that the condition on row 16 is a claw, and \cref{sec:list} releases two further state words, turns row 17 into a test against a list, gives the criterion that decides whether the claw and the list pay on other characteristics, and applies the list to 39-step SHA-512. \Cref{sec:tail} treats the uncontrolled steps 18 to 38, the event on which they are priced and its relation to the characteristic. \Cref{sec:attack} gives the attack and its complexity, \cref{sec:pinning} compares it with joint message-word pinning, \cref{sec:verification} states the premises and their verification, and \cref{sec:discussion} lists open problems. The appendices are supplementary material: proofs (\cref{app:proofs}), details of the implementation and the measurements (\cref{app:impl}), of the verification and reproduction (\cref{app:verif}), the characteristic, the SFS pairs and a worked example (\cref{app:tables}), and the provenance of the filings (\cref{app:provenance}).

\section{Preliminaries}\label{sec:prelim}

\subsection{Notation}
\begin{enumerate}
\item $\boxplus$ and $\boxminus$ denote addition and subtraction modulo $2^{32}$.
\item $\gg$, $\ggg$, $\oplus$, $\neg$, $\vee$ and $\wedge$ denote \emph{shift right}, \emph{rotate right}, \emph{exclusive or}, \emph{not}, \emph{or} and \emph{and}.
\item $x[i]$ denotes the $i$-th bit of $x$, and $x[0]$ is the least significant bit.
\item $\delta x$ denotes the modular difference, $\delta x = x' \boxminus x$, and $\nabla x$ the signed difference between $x'$ and $x$, in the notation of~\cite{LZL+26}: $\nabla x[i]$ is \texttt{n} for $(x[i],x'[i]) = (0,1)$, \texttt{u} for $(1,0)$, \texttt{=} for equal bits, and \texttt{0} or \texttt{1} for equal bits of that value.
\item $A_i$ and $E_i$ are the state words produced by step $i$; the chaining value is $\cv = (A_{-1},\dots,A_{-4},E_{-1},\dots,E_{-4})$. $W_i$ and $K_i$ are the expanded message word and the constant of step $i$.
\item $a_0 = A_0$ and $a_1 = A_1$ are the free state words and $u = W_1$; $D_8$ and $D_9$ are the message differences on $W_8$ and $W_9$; $g$ and $w_{17}$ are the published values of $W_{16}$ and $W_{17}$.
\item Hexadecimal constants are written in typewriter type.
\end{enumerate}

\subsection{Description of \sha{}}

\sha{} is a Merkle--Damg{\aa}rd hash function whose compression function, in Davies--Meyer mode, maps a 256-bit chaining value and a 512-bit message block to a 256-bit output~\cite{FIPS180-4}. The block is split into 16 words $W_0,\dots,W_{15}$, which the message expansion extends by
\[
W_i = \sig{1}(W_{i-2}) \boxplus W_{i-7} \boxplus \sig{0}(W_{i-15}) \boxplus W_{i-16}, \qquad i \ge 16,
\]
with $\sig{0}(x) = (x \ggg 7) \oplus (x \ggg 18) \oplus (x \gg 3)$ and $\sig{1}(x) = (x \ggg 17) \oplus (x \ggg 19) \oplus (x \gg 10)$. We use the alternative description of the state update in which only the words $A_i$ and $E_i$ produced by each step appear~\cite{MNS11,LZL+26}. Step $i$ computes
\begin{align*}
E_i &= A_{i-4} \boxplus E_{i-4} \boxplus \Sig{1}(E_{i-1}) \boxplus \IF(E_{i-1},E_{i-2},E_{i-3}) \boxplus K_i \boxplus W_i,\\
A_i &= E_i \boxminus A_{i-4} \boxplus \Sig{0}(A_{i-1}) \boxplus \MAJ(A_{i-1},A_{i-2},A_{i-3}),
\end{align*}
where $\IF(x,y,z) = (x \wedge y) \oplus (x \wedge z) \oplus z$, $\MAJ(x,y,z) = (x \wedge y) \oplus (x \wedge z) \oplus (y \wedge z)$, $\Sig{0}(x) = (x\ggg 2)\oplus(x\ggg 13)\oplus(x\ggg 22)$ and $\Sig{1}(x)=(x\ggg 6)\oplus(x\ggg 11)\oplus(x\ggg 25)$. We write $T_i = E_i \boxminus A_{i-4}$ for the part of step $i$ that does not depend on $A_{i-4}$, so that $A_i = T_i \boxplus \Sig{0}(A_{i-1}) \boxplus \MAJ(A_{i-1},A_{i-2},A_{i-3})$. The state after step $i$ is determined by $A_{i-7},\dots,A_i$. After the last step the chaining value is added to the state (the feed-forward). Each step can be inverted given its output: the state before step $i$ and $A_i$ (or $E_i$) determine $W_i$. The target of this paper is \sha{} with the compression function reduced to steps $0,\dots,38$ in every block, the standard IV, standard padding, the feed-forward and the full 256-bit digest.

\subsection{Collision attacks in two blocks}

A \emph{semi-free-start} (SFS) collision is a collision of the compression function from a chaining value that the attacker chooses and both messages share; a \emph{collision} must start from the IV. The attacks on 31 to 38 steps~\cite{MNS13,LLW+24b,LLWS26,LZL+26} turn an SFS-type characteristic into a collision in two blocks. A random first block $M_0$ supplies a chaining value $\cv$, and the second block carries the message difference. In the memory-efficient framework of~\cite{LLW+24b,LLWS26}, as used for 38 steps in~\cite[Sect.~3.1]{LZL+26}:
\begin{description}
\item[Step 1.] Find, once, a solution of the dense part of the characteristic: the message words $W_8,\dots,W_{13}$ and the state words $A_0,\dots,A_{13}$ and $E_4,\dots,E_{13}$.
\item[Step 2.] For each random $M_0$, connect $\cv$ to that solution by computing $W_0,\dots,W_6$ backwards, after checking the few conditions this connection imposes.
\item[Step 3.] Use the remaining freedom in $W_{14}, W_{15}$ and satisfy every remaining condition on $(A_i,E_i,W_i)_{i\ge 16}$ by trial; on failure, return to Step 2.
\end{description}
At 39 steps, the connection of Step 2 holds with probability 1, because the shifted local collision imposes no conditions on $W_5, W_6, W_7, E_1, E_2, E_3$~\cite[Sect.~4.3]{LZL+26}. The cost is then dominated by Step 3, which faces 168 conditions.

\subsection{Claws and parallel collision search}

A \emph{claw} between two functions $f_0, f_1$ on $n$-bit words is a pair $(x,y)$ with $f_0(x)=f_1(y)$. A claw can be found as a collision of the single map $F$ that applies $f_0$ on one half of the domain and $f_1$ on the other: a collision whose two inputs lie in different halves is a claw. The parallel collision search of van Oorschot and Wiener~\cite{vOW99} finds collisions of an $n$-bit map with walks that stop at \emph{distinguished points} (here, values whose low 7 bits are zero) and a table of their endpoints; two walks that reach the same distinguished point are re-walked to locate the collision. One collision costs about $2^{n/2}$ evaluations, and when many collisions are harvested from the same map with a table of $w$ entries, the amortized cost per collision falls to a small multiple of $\sqrt{2^n/w}$. For $n=32$ and a table that fits in one MiB, the attack described below measures about $2^{11.1}$ evaluations per distinct claw.

\subsection{Cost models}\label{sec:costmodels}

Most of the attack's work consists of evaluations of a 32-bit map rather than compressions, so its cost in compression calls depends on how operations are priced. We use two accountings.
\begin{itemize}
\item \emph{Compression-equivalents}, our headline unit. The attack runs on a bit-sliced machine whose instructions are XOR, AND, OR and NOT of 256-bit registers, one 32-bit lane per bit position of 256 registers. Every operation of the attack is counted, and the count is divided by the number of operations the same machine needs for one 39-step compression with its message schedule, measured on a straight-line program of that compression: $205.6$ operations per lane~\cite{PR358}. This is an operation count, not a timing; the literature's figures are timings of scalar or SHA-NI compressions, so the two units agree only up to the relative speed of the bit-sliced machine and the timed one. On the machine of \cref{sec:pinning}, one compression-equivalent costs $0.33$ scalar compressions or $3.2$ compressions with the ARMv8 SHA-256 instructions. The generic birthday search priced the same way costs about $2^{128}$.
\item \emph{The board's model}~\cite{HashSmash}: a classical probabilistic word RAM with 256-bit words, in which one 39-step compression costs one unit and every other primitive operation (load, store, addition, bitwise operation, shift, comparison, branch, random word) costs $1/C$ units with $C=2812$. Total time includes preprocessing, the construction of advice, failed trials and success amplification; success probability must be at least $0.39$; memory is the peak over all phases and is penalized by one point per doubling above $2^{20}$ bytes. Since $2812/205.6 = 2^{3.77}$, a cost in this model is $3.77$ bits below the same cost in compression-equivalents. The best generic search filed on the board costs $2^{124.3}$ in this model~\cite{PR358}.
\end{itemize}

\section{The 39-Step Characteristic and Its Free Words}\label{sec:transplant}

\subsection{The published characteristic and pair}

Table~21 of~\cite{LZL+26} (\cref{tab:char}) is a 39-step signed-difference characteristic of \sha{} with message differences in
\[
(W_8, W_9, W_{10}, W_{11}, W_{12}, W_{16}, W_{17}, W_{24}, W_{26}),
\]
the 38-step local collision shifted by one step~\cite[Sect.~4.3]{LZL+26}. Table~23 of~\cite{LZL+26} (\cref{tab:pair}) is an SFS colliding pair for it: a chaining value $\cv_{\mathrm{pub}}$ and two blocks $M'$, $M$, 160 bytes in all. In the signed differences of Table~21, $x$ is the block $M'$ and $x'$ the block $M$. We call the message that follows $M'$ the \emph{first member} and the one that follows $M$ the \emph{partner}, and take differences as partner minus first member: $D_8 = \delta W_8 = \hex{40000000}$ and $D_9 = \delta W_9 = \hex{f7801000}$.

We call rows 2 to 15 the \emph{dense part}: in these rows, the $A$ and $E$ words of both members and the words $W_{10},\dots,W_{15}$ are fixed by the published pair. Rows 16 to 38 are the \emph{tail}, where the published pair is one solution among many. Exhaustive enumeration over all $2^{32}$ values shows that exactly two values of $W_{16}$ complete row 16 from the dense part, $G_{16}=\{\hex{bca1c121},\hex{bca1c321}\}$, and that only $g=\hex{bca1c321}$ admits a row-17 word, of which there is exactly one, $w_{17}=\hex{71e9ba98}$.

\subsection{Two free state words}\label{sec:free}

In the framework of~\cref{sec:prelim}, the dense solution includes the first state words, and the only quantity that changes between trials is the chaining value. At 39 steps, this fixes more than necessary. The local collision starts at $W_8$, so $W_5, W_6, W_7$ and $E_1, E_2, E_3$ carry no conditions~\cite[Sect.~4.3]{LZL+26}, and neither do $E_4$ and $E_5$, the words through which $A_0$ and $A_1$ enter the dense part. We therefore pin only the state from step 2 onward, namely $A_2,\dots,A_5$ and $E_6,\dots,E_9$ (which, together with $W_{10},\dots,W_{15}$, fix rows 2 to 15), and leave $a_0=A_0$ and $a_1=A_1$ free. Each first block then yields $2^{64}$ candidate second blocks instead of a handful.

At 38 steps the local collision starts at $W_7$, and freeing $A_1$ as well costs conditions on $E_5$ and $W_7$ (\cref{sec:criterion}).

\subsection{The transplant}

The proofs of the statements of this section are in \cref{app:proofs}.

\begin{lemma}[Transplant]\label{lem:transplant}
Fix $A_2,\dots,A_5$, $E_6,\dots,E_9$ and $W_{10},\dots,W_{15}$ to the published values of the first member. For every chaining value $\cv$ and every $(a_0,a_1)$, steps 0 to 9 determine $W_0,\dots,W_9$ uniquely, and the first member then follows the published trajectory through step 15. Writing
\begin{align*}
\tau_0 &= \Sig{0}(A_{-1}) \boxplus \MAJ(A_{-1},A_{-2},A_{-3}), \qquad E_0 = a_0 \boxplus A_{-4} \boxminus \tau_0,\\
k_0 &= \boxminus \tau_0 \boxminus E_{-4} \boxminus \Sig{1}(E_{-1}) \boxminus \IF(E_{-1},E_{-2},E_{-3}) \boxminus K_0,
\end{align*}
we have $W_0 = a_0 \boxplus k_0$; with $u = W_1$,
\[
a_1 = u \boxplus \Sig{0}(a_0) \boxplus \MAJ(a_0,A_{-1},A_{-2}) \boxplus \Sig{1}(E_0) \boxplus \IF(E_0,E_{-1},E_{-2}) \boxplus E_{-3} \boxplus K_1,
\]
so that $a_1 \leftrightarrow u$ is a bijection for fixed $(\cv,a_0)$; and $W_9 = c_9 \boxminus a_1$ with the constant $c_9 = W_9^{\mathrm{pub}} \boxplus A_1^{\mathrm{pub}}$. The partner takes $W_8 \boxplus D_8$, $W_9 \boxplus D_9$ and the published $W_{10},\dots,W_{15}$ of $M$ and follows its published trajectory through step 15 in the same way.
\end{lemma}

\begin{theorem}[Bijection]\label{thm:bijection}
For the pinned dense part, the map
\[
\Phi:\ (\cv, a_0, a_1) \longmapsto (W_8, W_9, W_{16}, W_{17}, \dots, W_{23})
\]
is a bijection on $(\mathbb{Z}/2^{32})^{10}$, triangular in five stages:
\begin{enumerate}
\item $a_1 \leftrightarrow W_9$ is affine of unit slope and independent of $a_0$ and $\cv$;
\item given $a_1$, $a_0 \leftrightarrow W_8$ is affine of unit slope and independent of $\cv$;
\item given $(a_0,a_1)$, $(A_{-1},\dots,A_{-4}) \leftrightarrow (W_7,\dots,W_4)$;
\item given $(a_0,a_1,A_{-1},\dots,A_{-4})$, $(E_{-1},\dots,E_{-4}) \leftrightarrow (W_3,\dots,W_0)$;
\item given $W_8,\dots,W_{15}$, the expansion maps $(W_0,\dots,W_7)$ to $(W_{16},\dots,W_{23})$ unitriangularly.
\end{enumerate}
\end{theorem}

Each stage is triangular with unit slopes and is solved one word at a time; the proof is in \cref{app:proofs}.

Two consequences follow. First, over uniform $(a_0,a_1)$ the pair $(W_8,W_9)$ is exactly uniform for every $\cv$, so the side conditions below have exactly their counting probabilities. Second, the words $W_{18},\dots,W_{23}$ that drive the tail are a bijective image of free inputs, so the tail can be sampled on uniform draws and every sample inverted to an actual pair (\cref{sec:tail}).

\subsection{Rejoin conditions}

After the transplant, both members follow the published trail through row 15. Three conditions must hold for the differences to propagate through the message expansion as published; the published pair satisfies them, but a transplanted pair need not.

\begin{lemma}[Side conditions]\label{lem:side}
Let $\mathcal{F}_8 = \{w : \sig{0}(w+D_8) \boxminus \sig{0}(w) = \hex{087ff000}\}$ and $\mathcal{F}_9 = \{w : \sig{0}(w+D_9) \boxminus \sig{0}(w) = \hex{03010000}\}$. Membership in $\mathcal{F}_8$ depends only on the 14 bits of $w$ selected by $\hex{CC19E60C}$, of whose $2^{14}$ values exactly 1152 pass; membership in $\mathcal{F}_9$ depends only on the 17 bits selected by $\hex{F9983C0F}$, of whose $2^{17}$ values exactly 2080 pass. Hence $p_8 = 9/128 = 2^{-3.830}$, $p_9 = 65/4096 = 2^{-5.978}$, and under \cref{thm:bijection} the two events are independent, with joint probability $p_8 p_9 = 585/2^{19} = 2^{-9.8077}$.
\end{lemma}

In \cref{tab:char}, $\delta W_{16} = \boxminus\hex{087ff000}$, $\delta W_{17} = -2^{30}$, $\delta W_{23} = 0$ and $\delta W_{24} = \hex{03010000}$. In $W_{23} = \sig{1}(W_{21}) \boxplus W_{16} \boxplus \sig{0}(W_8) \boxplus W_7$ the $\sig{0}$ term must contribute $\hex{087ff000}$ for $W_{23}$ to be difference-free, which is the condition $W_8 \in \mathcal{F}_8$; in $W_{24} = \sig{1}(W_{22}) \boxplus W_{17} \boxplus \sig{0}(W_9) \boxplus W_8$ the differences of $W_{17}$ and $W_8$ cancel, so the $\sig{0}$ term must contribute $\hex{03010000}$, which is the condition $W_9 \in \mathcal{F}_9$. An exhaustive enumeration of both sets over all $2^{32}$ words agrees with these counts. The third condition is $W_{17} = w_{17}$, the only row-17 word that exists for $g$; for a uniform $W_{17}$ it has probability $2^{-32}$.

\section{Row 16 as a Claw}\label{sec:claw}

\subsection{Separation of the row-16 condition}

The expansion gives $W_{16} = \sig{1}(W_{14}) \boxplus W_9 \boxplus \sig{0}(W_1) \boxplus W_0$. Under the transplant, $W_{14}$ is pinned, $W_0$ is affine in $a_0$, $W_9$ is affine in $a_1$, and $a_1$ can be traded for $u = W_1$. The condition $W_{16} = g$, a 32-bit event that the framework of~\cite{LZL+26} satisfies by trial, therefore separates into a condition on $a_0$ and one on $u$.

\begin{proposition}\label{prop:claw}
For a fixed chaining value, define the 32-bit map
\[
F(x) = \begin{cases} \Feven(x) & x[0] = 0,\\ \Fodd(x) = \sig{0}(x) \boxminus x & x[0] = 1,\end{cases}
\]
where
\[
\Feven(x) = \Sig{0}(x) \boxplus \MAJ(x, A_{-1}, A_{-2}) \boxplus \Sig{1}(E_0(x)) \boxplus \IF(E_0(x), E_{-1}, E_{-2}) \boxminus x \boxplus c,
\]
with $E_0(x) = x \boxplus A_{-4} \boxminus \tau_0$ and $c = g \boxminus \sig{1}(W_{14}) \boxminus c_9 \boxminus k_0 \boxplus E_{-3} \boxplus K_1$ as in \cref{lem:transplant}. Then for every $(a_0, u)$,
\[
W_{16} \boxminus g = \bigl(\sig{0}(u) \boxminus u\bigr) \boxminus \Feven(a_0),
\]
so that, with $a_0[0] = 0$ and $u[0] = 1$, $W_{16} = g$ if and only if $F(a_0) = F(u)$.
\end{proposition}

\begin{proof}
Substitute $W_0 = a_0 \boxplus k_0$, $W_1 = u$ and $W_9 = c_9 \boxminus a_1$ from \cref{lem:transplant} into the expansion, and then the expression of $a_1$ in terms of $(a_0,u)$.
\end{proof}

Thus $W_{16} = g$ exactly when $(a_0, u)$ is a claw between the even and the odd branches of a single map $F$ on 32-bit words. The branch $\Feven$ depends on the chaining value through six constants of the first block, so the map changes with every first block. The restriction to even $a_0$ and odd $u$ costs a constant factor in the number of claws and nothing in exactness; its effect on the later events is treated in \cref{sec:parity}.

With only $a_0$ free, the row-16 condition is a 32-bit equation in one variable for each chaining value, and there is no claw; \cref{sec:criterion} gives the price of freeing $a_1$ on other characteristics.

\subsection{Finding claws}\label{sec:walk}

Claws are found by parallel collision search on $F$~\cite{vOW99}. One \emph{window} is the work done on one first block: one compression for $\cv$, the constants of $F$, and $2^{26}$ evaluations of $F$. Trails of 256 lanes advance until they reach a distinguished point, a value whose low 7 bits are zero, which is stored in a table that fits in one MiB; two trails that reach the same distinguished point are re-walked to their first common value, whose two predecessors form a claw if their parities differ. The walk evaluates $F$ as a straight-line bit-sliced program generated for each first block, at 2.8 to 3.7 operations per lane against 205.6 for a compression, and every operation is counted. In the attack of \cref{sec:attack}, a window yields about 57{,}000 distinct claws at about 1{,}900 evaluations of $F$ each, and a distinct claw costs $40.0$ compression-equivalents ($2.9$ in the board's model). \Cref{app:walk} gives the parameters of the walk, its measured yields and its operation counts.

\subsection{Parity and the fibre of the tail}\label{sec:parity}

Every claw has $a_0$ even and $u$ odd. Under \cref{thm:bijection} these become conditions on $(W_8, W_9, W_{16},\dots,W_{23})$. The condition on $a_0$ fixes bit 0 of $W_8$ as a function of $W_9$; bit 0 is not among the active bits of $\mathcal{F}_8$, so it selects exactly half of $\mathcal{F}_8$. For $u$, the expansion read downwards from $W_{17} = w_{17}$ gives $u = C_1 \boxminus \sig{0}(W_2)$ with the odd constant $C_1 = w_{17} \boxminus \sig{1}(W_{15}) \boxminus W_{10} = \hex{f918ec03}$, and flipping bit 18 of $W_{18}$ flips the parity of $\sig{0}(W_2)$ without changing $(W_8,W_9,W_{19},\dots,W_{23})$. Hence, on the fibre of the tail, the event that $u$ is odd has probability exactly $1/2$, and the tail probability can be measured on the inputs that claws produce, the \emph{dual-parity fibre}.

\section{Four Free State Words: \texorpdfstring{$W_{17}$}{W17} over a List}\label{sec:list}

\begin{figure}[t]
\centering
\begin{tikzpicture}[x=0.42cm,y=0.55cm,font=\scriptsize,
  cell/.style={draw=black!55,inner sep=0pt,anchor=south west,minimum height=0.495cm},
  dark/.style={cell,fill=black!75,text=white},
  mid/.style={cell,fill=black!45,text=white},
  cvc/.style={cell,fill=black!22},
  adv/.style={cell,fill=black!7},
  side/.style={cell,pattern=north east lines,pattern color=black!45},
  open/.style={cell}]
\newcommand{\seg}[5]{\node[#1,minimum width={(#3-#2+1)*0.42cm}] at (#2+4,#4) {#5};}
\newcommand{\tail}[2]{\node[open,minimum width=4.5*0.42cm] at (22,#1) {#2};}
\foreach \y/\l in {2.2/A_i, 1.1/E_i, 0/W_i} \node[anchor=east] at (-0.2,\y+0.45) {$\l$};
\seg{cvc}{-4}{-1}{2.2}{$\cv$}
\seg{dark}{0}{0}{2.2}{$a_0$}\seg{dark}{1}{1}{2.2}{$a_1$}\seg{mid}{2}{2}{2.2}{$a_2$}\seg{mid}{3}{3}{2.2}{$a_3$}
\seg{adv}{4}{15}{2.2}{$A_4,\dots,A_7$ pinned}
\seg{dark}{16}{16}{2.2}{}\seg{mid}{17}{17}{2.2}{}\tail{2.2}{$2^{-74.5}$}
\seg{cvc}{-4}{-1}{1.1}{$\cv$}
\seg{open}{0}{7}{1.1}{from $\cv$ and $a_0,\dots,a_3$}
\seg{adv}{8}{15}{1.1}{$E_8,\dots,E_{11}$ pinned}
\seg{dark}{16}{16}{1.1}{}\seg{mid}{17}{17}{1.1}{}\tail{1.1}{}
\seg{open}{0}{7}{0}{from $\cv$ and $a_0,\dots,a_3$}
\seg{side}{8}{11}{0}{}
\seg{adv}{12}{15}{0}{pinned}
\seg{dark}{16}{16}{0}{}\seg{mid}{17}{17}{0}{}\tail{0}{expansion}
\foreach \x/\t in {2/first block, 6/free, 14/advice, 21/built, 24.25/by probability} \node at (\x,3.55) {\t};
\foreach \i in {-4,0,4,8,12,16,17} \node at (\i+4.5,-0.45) {$\i$};
\node at (24.25,-0.45) {18 to 38};
\node[anchor=east] at (-0.2,-0.45) {step};
\end{tikzpicture}
\caption{The second block of the attack with four free state words. The chaining value comes from the first block, $a_0,\dots,a_3$ are free, $A_4,\dots,A_7$, $E_8,\dots,E_{11}$ and $W_{12},\dots,W_{15}$ are pinned to the advice, and the rest of steps 0 to 15 follows. Rows 16 and 17 are met by construction: row 16 by a claw in $a_0$ and $a_1$ (dark), row 17 by the list over $a_2$ and $a_3$ on a fibre of $h$ (medium). Hatched: $W_8,\dots,W_{11}$, which carry the side conditions. Steps 18 to 38 are paid by probability.}\label{fig:block}
\end{figure}
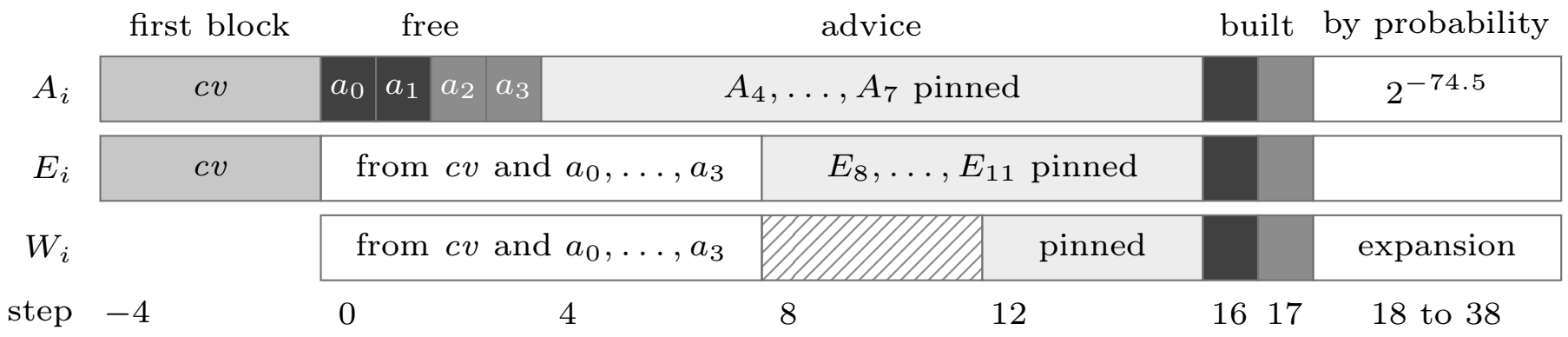

With two free words, the condition $W_{17}=w_{17}$ is a trial that succeeds with probability $2^{-32}$ per claw. Releasing two more state words, $a_2 = A_2$ and $a_3 = A_3$, on a set where the claw is unchanged lets one claw test $W_{17}$ against several million second blocks. This section gives the construction as it stands in the reference implementation~\cite{PR1809}; \cref{sec:attack} gives the attack and its cost, and \cref{tab:levers} in \cref{app:provenance} the steps by which it was reached. \Cref{fig:block} shows the second block of the attack.

\subsection{The transplant with four free words}

The advice is a 160-byte SFS pair on the characteristic of \cref{tab:char}, re-solved for a good class of $\sig{0}(W_{11})$ (\cref{sec:class}); \cref{tab:pair2} lists it. Its modular differences and its row-16 and row-17 words $g$, $w_{17}$ are those of the published pair. We pin $A_4,\dots,A_7$, $E_8,\dots,E_{11}$ and $W_{12},\dots,W_{15}$ to the advice and leave $(a_0,a_1,a_2,a_3)$ free.

\begin{lemma}[Transplant, four free words]\label{lem:four}
For every chaining value and every $(a_0,a_1,a_2,a_3)$, steps 0 to 11 determine $W_0,\dots,W_{11}$ uniquely, and the first member then follows the advice through step 15. Write $E_7 = a_3 \boxplus c_7$ and $E_6 = a_2 \boxplus t(a_3)$ with
\[
c_7 = A_7 \boxminus \Sig{0}(A_6) \boxminus \MAJ(A_6,A_5,A_4), \quad t(a_3) = A_6 \boxminus \Sig{0}(A_5) \boxminus \MAJ(A_5,A_4,a_3),
\]
and
\[
h(a_2,a_3) = \MAJ(A_4,a_3,a_2) \boxminus \IF(E_8,E_7,E_6).
\]
Then
\begin{enumerate}
\item $W_0$ and $W_1$ do not depend on $(a_2,a_3)$ and are those of \cref{lem:transplant};
\item $W_2 = a_2 \boxplus w$, where $w$ depends on $(\cv,a_0,a_1)$ only;
\item $W_9 = c_9' \boxminus a_1 \boxplus h(a_2,a_3)$, with the constant $c_9' = E_9 \boxminus 2A_5 \boxplus \Sig{0}(A_4) \boxminus \Sig{1}(E_8) \boxminus K_9$;
\item $W_{10} = \kappa(a_3) \boxminus a_2$, with $\kappa(a_3) = E_{10} \boxminus A_6 \boxminus t(a_3) \boxminus \Sig{1}(E_9) \boxminus \IF(E_9,E_8,E_7) \boxminus K_{10}$;
\item $W_{11}$ depends on $a_3$ only, and $W_8$ depends on $(a_0,a_1,a_2,a_3)$ and not on $\cv$.
\end{enumerate}
\end{lemma}

The proof is in \cref{app:proofs}.

\subsection{The list on a fibre of \texorpdfstring{$h$}{h}}

By \cref{lem:four}, the pair $(a_2,a_3)$ affects $W_9$, and through it $W_{16}$, only through the value $h(a_2,a_3)$.

\begin{definition}
For a 32-bit value $\eta$, the \emph{list} $L_\eta$ is the set of pairs $(a_2,a_3)$ on the fibre $h(a_2,a_3)=\eta$ for which both members of the transplanted pair follow the advice trail through step 15.
\end{definition}

On $L_\eta$, the word $W_9 = c_9' \boxminus a_1 \boxplus \eta$ takes one value for each $a_1$, so $\mathcal{F}_9$ remains a test on the claw, and \cref{prop:claw} holds with $c_9$ replaced by $c_9' \boxplus \eta$: a claw of the shifted map gives $W_{16}=g$ for every entry of the list. The fibre value $\eta$ is a parameter. At the value of $h$ at the advice's own words, $\eta_0 = \hex{36faf0de}$, the list has 821{,}520 entries; at $\eta = \hex{3cc9eebe}$, the fibre of the reference implementation, it has 9{,}701{,}592 entries on 728 values of $a_3$.

\paragraph{The list by rule.} Let $\mathcal{F}_i = \{w : \sig{0}(w \boxplus D_i) \boxminus \sig{0}(w) = T_i\}$ for $i = 10, 11$, with the advice's differences
\begin{align*}
(D_{10},T_{10}) &= (\hex{c0000000},\hex{087ff000}),\\
(D_{11},T_{11}) &= (\hex{0601dcd4},\hex{1fff41a0}).
\end{align*}
The reference implementation defines $L_{\hex{3cc9eebe}}$ as the set of $(a_2,a_3)$ with $h(a_2,a_3) = \eta$ such that $E_7$ differs from the advice's value only inside the 14 bits selected by \hex{f74400da}, $W_{11}\in\mathcal{F}_{11}$, $W_{10}\in\mathcal{F}_{10}$ and $E_6[30]=E_7[30]$,. The two conditions on $W_{10}$ and $W_{11}$ keep the $\sig{0}$ differences that enter $W_{25}$ and $W_{26}$ at the advice's values, so the list stays in the advice's class (\cref{sec:class}). For each $a_3$, the fibre equation is solved for $a_2$ bit by bit along its two carry chains, so the list is generated without being stored. The rule returns the same set as an enumeration over all $2^{32}$ values of $a_3$ with the trail of both members checked on every entry, which we repeated independently (\cref{app:impl}); the agreement depends on the fibre (\cref{sec:discussion}).

\subsection{\texorpdfstring{$W_{17}$}{W17} as an equation in \texorpdfstring{$a_2$}{a2}}

Substituting \cref{lem:four} into $W_{17} = \sig{1}(W_{15}) \boxplus W_{10} \boxplus \sig{0}(W_2) \boxplus W_1$ and writing $y = a_2 \boxplus w$,
\[
W_{17} = \sig{1}(W_{15}) \boxplus \kappa(a_3) \boxplus w \boxplus u \boxplus \bigl(\sig{0}(y) \boxminus y\bigr).
\]
So $W_{17} = w_{17}$ if and only if
\begin{equation}\label{eq:w17}
\Fodd(y) = r, \qquad r = w_{17} \boxminus \sig{1}(W_{15}) \boxminus \kappa(a_3) \boxminus w \boxminus u,
\end{equation}
an equation in the odd branch of the map $F$ of \cref{prop:claw}. Rows 16 and 17 are both equations in $x \mapsto \sig{0}(x) \boxminus x$. Entries with the same \emph{key} $(a_2,\kappa(a_3))$ satisfy \cref{eq:w17} simultaneously and are called \emph{mates}; they differ in $a_3$, hence in $W_3,\dots,W_8$, $W_{11}$ and every tail word. The list $L_{\hex{3cc9eebe}}$ has 3{,}462{,}264 keys in 200 groups of equal $\kappa$.

\paragraph{The pass.} In $\sig{0}(y) = y \boxplus r$, bit $i$ of the left side is $y[i+7] \oplus y[i+18] \oplus y[i+3]$ for $i \le 13$, so the equation's bits 0 to 13 give $y[18],\dots,y[31]$ from the low 18 bits of $y$, and bits 14 to 31 remain as 18 checks. The test therefore runs over the distinct values of $a_2 \bmod 2^{18}$ in a group and not over its keys. The list $L_{\hex{3cc9eebe}}$ has 62{,}400 such prefixes. They are evaluated 256 at a time in bit-sliced form, at most 143{,}633 operations per claw that passes $\mathcal{F}_9$, that is 699 compression-equivalents for 9.7 million entries. From bit 22 of the equation on, a block of 256 prefixes is abandoned as soon as none of its lanes survives, which brings the mean to about 126{,}800 operations. A surviving lane is tested in scalar code; at a key hit every mate with $W_8 \in \mathcal{F}_8$ is built and compressed. The condition on $W_8$ is not part of the bit-sliced pass: since $W_{17}$ passes with probability $2^{-32}$, $\mathcal{F}_8$ is tested at hits only, which removes 55\% of the counted cost of the pass.

\subsection{The claw pays when the local collision starts at \texorpdfstring{$W_7$}{W7} or later}\label{sec:criterion}

\Cref{prop:claw,lem:four} use only the step function and the message expansion, not the characteristic, and with the functions of SHA-512 they hold verbatim on 64-bit words. We checked both identities on 2{,}000 random pinned states for each word size, and checked with the same test that $W_{17}$ is not additively separable in $(a_0,a_1)$, so row 16 is the only row that becomes a claw. On another characteristic the construction is therefore always available, and the characteristic sets only its price. By the two relations at the start of the proof of \cref{thm:bijection}, a free word $A_k$ enters only $A_k$, $E_k,\dots,E_{k+4}$ and $W_k,\dots,W_{k+8}$. Freeing $a_0$ and $a_1$ costs the conditions on $E_0,\dots,E_5$ and the side conditions of the message differences among $W_0,\dots,W_9$; freeing $a_2$ and $a_3$ costs in addition those on $E_6$, $E_7$, $W_{10}$ and $W_{11}$.

Let $p_{01}$ be the probability that, for a uniform chaining value and uniform $(a_0,a_1)$, the transplanted pair keeps the differential of a published SFS pair through step 15: the second member reaches the pair's state after step 9, and the $\sig{0}$ differences of $W_0,\dots,W_9$ are the pair's. For the characteristic of this paper $p_{01} = p_8p_9$ (\cref{lem:side}). \Cref{tab:criterion} gives $p_{01}$ for every published characteristic of 36 to 39 steps, measured on its pair (\cref{app:price}). In the framework of~\cite{LZL+26}, row 16 costs $2^{c_{16}}$ trials, where $c_{16}$ is the number of its conditions; the claw replaces them by about $2^{5.3}/p_{01}$ compression-equivalents (\cref{sec:walk}). On SHA-512, $\Fodd$ is inverted by guessing 8 bits at about 3 compressions~\cite{GWZW26}, and the inversion takes the place of the claw.

\begin{table}[t]
\centering
\caption{The price of freeing $A_0$ and $A_1$ on the published characteristics, measured on their SFS pairs with $2^{28}$ draws ($-\log_2$; bounds at 95\%). $c_{16}$: the conditions on row 16, counted from the published tables, or taken from~\cite{GWZW26} for the 38-step characteristics of SHA-256. Gain: $c_{16}$ less the price and the cost of one solution of row 16, which is $2^{5.3}$ compression-equivalents for the claw on SHA-256 and $2^{1.5}$ to $2^{1.7}$ compressions for the inversion on SHA-512~\cite{GWZW26}; read it to about a bit.}\label{tab:criterion}
\setlength{\tabcolsep}{5pt}
\begin{tabular}{@{}llcrrr@{}}
\toprule
Hash & Characteristic & First $\delta W$ & $-\log_2 p_{01}$ & $c_{16}$ & Gain \\
\midrule
SHA-256 & 39 steps~\cite{LZL+26} & $W_8$ & $9.8$ & $31$ & $15.9$ \\
SHA-512 & 39 steps~\cite{LZL+26} & $W_8$ & $9.0$ & $53$ & $42.5$ \\
\addlinespace
SHA-256 & 38 steps~\cite{LZL+26} & $W_7$ & $19.0$ & $29$ & $4.7$ \\
SHA-256 & 38 steps~\cite{GWZW26} & $W_7$ & $17.6$ & $25$ & $2.1$ \\
SHA-512 & 38 steps~\cite{LZL+26} & $W_7$ & $13.0$ & $41$ & $26.3$ \\
\addlinespace
SHA-256 & 37 steps~\cite{LZL+26} & $W_6$ & $>26.6$ & $20$ & $<-11.9$ \\
SHA-256 & 37 steps~\cite{ZMLM26} & $W_6$ & $>26.6$ & $5$ & $<-26.9$ \\
SHA-512 & 37 steps~\cite{LZL+26} & $W_6$ & $>26.6$ & $24$ & $<-4.1$ \\
SHA-256 & 36 steps~\cite{LZL+26} & $W_5$ & $>26.6$ & $15$ & $<-16.9$ \\
SHA-256 & 36 steps~\cite{ZMLM26} & $W_5$ & $>26.6$ & $9$ & $<-22.9$ \\
SHA-512 & 36 steps~\cite{LZL+26} & $W_5$ & $>26.6$ & $18$ & $<-10.1$ \\
\bottomrule
\end{tabular}
\end{table}

The table splits at the start of the local collision. When it starts at $W_8$, the reach of $a_0$ and $a_1$ carries no condition on the state for either word size, and the price is that of the side conditions of $W_8$ and $W_9$. When it starts at $W_7$, the reach takes in the condition of the $\IF$ of step 8 on $E_5$ and the side condition of $W_7$; the price, 13 to 19 bits, stays below the 25 to 41 conditions of row 16, and on SHA-256 the claw still gains 2 to 5 bits. When it starts at $W_6$ or earlier, $E_4$ and $E_5$ lie in the dense part, and the price, above 26.6 bits, exceeds the 5 to 24 conditions of row 16.

The list follows the same rule. On the published 39-step pair, freeing $a_2$ and $a_3$ costs $2^{-28.0}$: $2^{-8.0}$ for the state through step 11 and $2^{-19.8}$ for the side conditions of $W_{10}$ and $W_{11}$. That leaves about $2^{36}$ pairs $(a_2,a_3)$, of which the fibre of the attack holds $2^{23.2}$ on the advice. On every characteristic of 38 steps or fewer, the state part alone has probability below $2^{-26.6}$ (no hit in $2^{28}$ draws), at least 18.6 bits below; scaled in proportion, a fibre would hold a few dozen entries at most. On 39-step SHA-512 the state part costs $2^{-21}$, and given it the side conditions cost $2^{-13.0}$: the two parts are correlated, the price is $2^{-34.0}$, and about $2^{94}$ pairs remain. \Cref{sec:sha512} builds the list on this case.

\subsection{On 39-step SHA-512, a bitwise search of the list gains 29 bits}\label{sec:sha512}

On 39-step SHA-512, single-word pinning~\cite{GWZW26} solves row 16 by inverting $\Fodd$ at a chosen target $W_{16} = \hex{e90725beef8b6d13}$ and leaves row 17 to chance. With the states through step 16 fixed, row 17 is a set of conditions on $W_{17}$ alone, met by a set $S_{17}$ of $1{,}404{,}469{,}248 = 2^{30.39}$ values~\cite{GWZW26}; with $\mathcal{F}_8$, a solution of row 16 passes with probability $2^{-36.61}$, and the attack costs $2^{131.3}$. \Cref{lem:four} holds on 64-bit words, so the list carries over, and on the list row 17 becomes a search instead of a trial.

\paragraph{The list.} Before step 16, $a_2$ and $a_3$ meet the differential only in the $\IF$ differences of steps 9 and 10. Step 9 requires $E_6[63] \ne E_7[63]$; step 10 depends on $E_7$ only at the 20 bits where $E_9$ carries a difference, and fixes them. With $W_{10}\in\mathcal{F}_{10}$, three parities of $W_{10}$, and $W_{11}\in\mathcal{F}_{11}$, this is the whole rule, at the price $2^{-34.0}$ of \cref{tab:price}. Fix a 64-bit fibre value $\eta$ and $a_3$, and call the entries with this $a_3$ a \emph{class}. Its values of $a_2$ are the words accepted by an automaton of 64 states that reads $a_2$ from bit 0 and tracks the carries of $E_6 = a_2 \boxplus t(a_3)$, of $h$ and of $W_{10} = \kappa(a_3) \boxminus a_2$, and the three parities. A class holds $2^{19}$ to $2^{38.4}$ entries, and one fibre holds at least $2^{46.1}$ (\cref{app:sha512}).

\paragraph{The search.} By \cref{lem:four}, $W_{17} = \sig{0}(a_2 \boxplus w) \boxplus (\kappa(a_3) \boxminus a_2) \boxplus k_1$, where $w$ and $k_1$ depend on $(\cv,a_0,a_1)$ only. On 64-bit words, bit $j \le 55$ of $\sig{0}(y)$ reads $y[j+1]$, $y[j+7]$ and $y[j+8]$, so once $a_2$ is fixed on bits 0 to $k$, $W_{17}$ is known on bits 0 to $k-8$. The conditions of row 17 are bit conditions on $W_{17}$, $E_{17}$ and $A_{17}$, and at fixed states the last two are $W_{17}$ plus constants. A depth-first search over the bits of $a_2$ along the live transitions of the class's automaton therefore checks row 17 at bit $k-8$ as soon as that bit is known, and row 17 in full and $\mathcal{F}_8$ at the leaves. It returns exactly the entries of the class that pass both, without visiting the others. On SHA-256 the list meets the single value $w_{17}$ of \cref{eq:w17}, and the pass works on 18-bit prefixes; on SHA-512, $S_{17}$ has $2^{30.4}$ values and the search runs 8 bits behind.

\paragraph{Cost.} \Cref{tab:sha512} gives the search on three classes, timed on one core of the machine of \cref{sec:pinning} in scalar 39-step SHA-512 compressions. A solution of row 16 that passes $\mathcal{F}_9$ costs $C_{16} = 2^{6.28}$, the inversion included. On class A the search costs $2^{11.69}$ and yields $0.2145 \pm 0.0081$ hits over 73{,}728 searches, against $n \cdot 2^{-36.61} = 0.219$ expected, so a hit costs $2^{13.92}$ ($2^{13.95}$ at the measured rate). Single-word pinning, timed with the same code, pays $C_{16} \cdot 2^{36.61} = 2^{42.89}$ per hit: the search gains 29.0 bits. Testing the entries one by one would gain at most $\log_2(C_{16}/c_e) = 13.8$ bits, where $c_e = 2^{-7.48}$ is the cost of one test. All 32{,}392 hits of these runs passed a run of both members through step 17, and 8{,}188 further hits passed a separate implementation of SHA-512 checked against Tables~24 and~25 of~\cite{LZL+26}.

\begin{table}[t]
\centering
\caption{The bitwise search of row 17 on three classes of 39-step SHA-512 ($\log_2$; scalar 39-step compressions, one core of an Apple M1 Pro). $n$: entries of the class. Nodes and search: per solution of row 16. Per hit: $C_{16}$ plus the search, over the expected hits per search $n \cdot 2^{-36.61}$, which the measured rates match (\cref{app:sha512}). Gain: over single-word pinning timed with the same code, $2^{42.89}$ per hit.}\label{tab:sha512}
\setlength{\tabcolsep}{6pt}
\begin{tabular}{@{}lrrrrr@{}}
\toprule
Class & $n$ & Nodes & Search & Per hit & Gain \\
\midrule
A & $34.42$ & $16.32$ & $11.69$ & $13.92$ & $28.97$ \\
B & $35.47$ & $17.57$ & $12.85$ & $14.01$ & $28.88$ \\
Largest found & $37.79$ & $20.74$ & $15.97$ & $14.80$ & $28.09$ \\
\bottomrule
\end{tabular}
\end{table}

\paragraph{Tail and estimate.} Steps 18 to 24 do not read the entry: $W_{10}$ first enters $W_{25}$, and $W_{11}$ enters $W_{18}$ next to $W_2$, which is uniform. On 32{,}768 survivors of step 24, entries of classes A and B pass steps 25 to 38 at the rate of the pair's own words within 0.08 bits at 95\%, and the tail from step 18 measures $86.96 \pm 0.03$ bits, against $87.21$ in~\cite{GWZW26}. A hit at $2^{13.92}$ with the tail of~\cite{GWZW26} gives about $2^{101.1}$ compressions per expected collision, against $2^{131.3}$ in~\cite{GWZW26} and $2^{130.1}$ for single-word pinning timed on our machine. The estimate composes measured factors under the premises of~\cite{GWZW26}. Hits cluster: 98\% of searches return none, and one returned 206. That leaves the expected cost unchanged, but no bound on the first collision like \cref{prop:success} has been derived, and no collision has been found. These measurements are not a filing on the board (\cref{app:provenance}).

\section{The Uncontrolled Part: Steps 18 to 38}\label{sec:tail}

For a claw that passes the filters and the test on $W_{17}$, both members follow the trail through row 17, and $W_{18},\dots,W_{23}$ are six fresh uniform words, the bijective image of $E_{18},\dots,E_{23}$. The \emph{tail event} is the event that the two 39-step compressions end in the same 256-bit state after step 38. Its probability $\ptail$ is the only quantity in the cost that is measured rather than counted.

\subsection{The priced event}\label{sec:carrier}

Let $U$ be the distribution in which $E_{16}$ and $E_{17}$ of the first member are pinned to their published values (which fixes $W_{16}=g$ and $W_{17}=w_{17}$), $W_8$ is uniform on $\mathcal{F}_8$ and $W_9$ on $\mathcal{F}_9$, $E_{18},\dots,E_{23}$ are uniform, and everything else is determined by the real step function and expansion. The cost is priced on
\[
Q = \Pr_U\bigl[\text{collision} \wedge K_P \mid a_0 \text{ even},\ u \text{ odd}\bigr],
\]
where the collision test is exact and $K_P$, the \emph{carrier}, is the following set of conditions on the two members.
\begin{description}
\item[Rows 16 and 17.] Fixed by the pinning of $E_{16}$ and $E_{17}$.
\item[Row 18.] $\nabla A_{18}$ is \texttt{u} at bit 30 and \texttt{=} elsewhere; $E_{18}$ and $W_{18}$ carry no difference; $E_{18}$ is 0 at bits 0, 8, 10, 11, 12, 15, 16, 23 and 31 and 1 at bit 17; $A_{18}[i]=A_{17}[i]$ for $i\in\{16,24,26,30\}$, $A_{18}[10]=A_{17}[21]$, $A_{18}[7]\ne A_{18}[19]$, $A_{18}[9]\ne A_{18}[18]$, $E_{18}[5]=E_{17}[5]$ and $E_{18}[19]=E_{17}[19]$.
\item[Rows 19 to 22.] The signed differences of $A_i$, $E_i$ and $W_i$ of \cref{tab:char}, and nothing else.
\item[Row 23.] No difference, and $E_{23}[30]=0$.
\item[Rows 24 to 38.] No condition.
\end{description}
The conditions at row 18 are those that Tables~21 and~22 of~\cite{LZL+26} place there, with two exceptions: $K_P$ does not check $E_{18}[24]=0$, and it takes the relation between bit 5 of $E_{17}$ and $E_{18}$ as an equality, which is how the published pair satisfies it (Table~22 prints an inequality). Since $\text{collision}\wedge K_P$ implies the collision, $Q \le \ptail$, and the cost uses $\ptail \ge Q$. The carrier relaxes the characteristic in two places: it does not check the conditions on the Boolean functions and $\Sig{1}$ at rows 19 to 22, which fix the signed differences of their outputs, and it checks nothing after row 23. \Cref{sec:gap} measures the gain from each relaxation.

\subsection{The estimator}\label{sec:smc}

The reference implementations estimate $Q$ with a sequential Monte Carlo sampler over rows 18 to 23, followed by plain Monte Carlo over rows 24 to 38.
\begin{itemize}
\item \emph{Stages 18 to 23.} A population of $N$ particles (here $N = 800$) holds states of both members through row $t-1$; at the start every particle is the pinned state through row 17. At stage $t$, each particle proposes $m_t$ candidate values of $E_t$ of the first member, uniform except that the $n_t$ bits that $K_P$ fixes at row $t$ are set to their values; $W_t$ and the words $A_t$, $E_t$ of both members follow. A candidate is accepted if it satisfies row $t$ of $K_P$ and an exact look-ahead: the modular differences of $A_{t+1}$, $E_{t+1}$ and $W_{t+1}$, and for $t \le 21$ of $W_{t+2}$, which the state after row $t$ already determines, must equal those of the characteristic. With $a_t$ accepted candidates, the stage factor is $\rho_t = 2^{-n_t} a_t/(N m_t)$, and the next population consists of $N$ draws with replacement, uniform among the accepted candidates. If $a_t = 0$, the estimate is 0.
\item \emph{Rows 24 to 38.} Each particle drives the same number of proposals. A proposal draws $W_8$ uniform on $\mathcal{F}_8$ and $W_9$ uniform on $\mathcal{F}_9$, computes $W_{24},\dots,W_{38}$ and the two compressions, and is a \emph{hit} if they end in the same state. With $h$ hits in $P$ proposals, the estimate is $\widehat{Q} = \rho_{18}\cdots\rho_{23}\cdot h/P$.
\end{itemize}
Every hit is inverted through \cref{thm:bijection} to a chaining value and checked by two 39-step compressions. These hits are SFS collisions: each starts from a chaining value derived from the sampled words, not from one produced by a first block from the IV; this is why they are cheap, and why none of them is a collision of the target.

\begin{proposition}\label{prop:smc}
$\mathbb{E}[\widehat{Q}] = \Pr_U[\text{collision} \wedge K_P]$.
\end{proposition}

The proof (\cref{app:proofs}) is that of the unbiasedness of the unnormalized Feynman--Kac particle estimator~\cite{DelMoral04}.

The estimator pools the four parity classes of $(a_0,u)$, which the proposals hit with probability $1/4$ each (\cref{sec:parity}), so it estimates $Q$ if the tail does not depend on the class. The class-specific estimator $4\rho_{18}\cdots\rho_{23}\,h_{01}/P$, where $h_{01}$ counts the hits with $a_0$ even and $u$ odd, is unbiased for $Q$ without that premise; \cref{sec:verification} reports both.

Our second estimator, used for the measurements this paper reports as its own, is fixed-effort splitting. Rows 16 and 17 are enumerated exhaustively, each of rows 18 to 22 draws a parent uniformly and a uniform $W_t$ until $N = 2000$ states pass the row, and $N\cdot T$ proposals with $T = 32{,}768$ draw $W_{23}$ and run the tail. Each stage rate is $N$ divided by a negative-binomial number of trials, so the product of the rates overstates the mean by a factor of at most $(1+1/(N-1))$ per stage, $0.004$ bits over five stages. The strict event of \cref{sec:gap} is measured with the same estimator, so the bias cancels in the comparison of the two events.

\subsection{Measurements}

For the published pair (\cref{tab:tail}), 14 organizer-run replicas of the reference implementation with two free words give 601 hits in 8{,}601{,}600 proposals, all verified as collisions, with mean $2^{-75.39}$ and one-sided 99\% Student lower bound $2^{-75.59}$ on the linear scale; that implementation declares $2^{-75.67}$. Our second estimator, on 40 seeds fixed before the run, gives $2^{-75.36}$ with a 99\% bound of $2^{-75.41}$.

For the entries of the list we ran the replica of the reference implementation, unmodified, under the law at which it prices the tail: one uniform listed entry per batch of 2{,}048 proposals, with $W_8$ and $W_9$ drawn per proposal. The 64 runs of 3{,}276{,}800 proposals each were pre-registered before any counted run (\cref{sec:verification}). They give 28{,}715 hits, all verified, and a mean of $2^{-74.45}$, with the 95\% interval $[-74.49,\ -74.41]$ and the one-sided 99.9\% bound $2^{-74.51}$, with the rate of each row in \cref{tab:carrierrows}. In these runs every pair that keeps the advice's differences through row 28 collides.

\begin{table}[t]
\centering
\caption{Measurements of the tail probability under the carrier $K_P$ ($\log_2$ of the mean on the linear scale, with one-sided lower bounds).}\label{tab:tail}
\setlength{\tabcolsep}{4pt}
\begin{tabularx}{\textwidth}{@{}l>{\raggedright\arraybackslash}Xrr@{}}
\toprule
Pair & Estimator and runs & Mean & Bound \\
\midrule
published & replica of~\cite{PR574}, 14 organizer runs & $-75.39$ & $-75.59$ (99\%) \\
published & the same replica, offline, $64\times 2^{28}$ samples & $-75.35$ & $-75.37$ (99\%) \\
published & second estimator, 40 seeds & $-75.36$ & $-75.41$ (99\%) \\
published & second estimator, 30 seeds (\cref{sec:gap}) & $-75.33$ & $-75.38$ (99\%) \\
good class \hex{2000c1a0} & second estimator, 4 seeds & $-74.44$ & \\
entries of the list & replica of~\cite{PR1809}, 14 organizer runs, pooled over classes of entries & $-74.46$ & $-74.66$ (99.9\%) \\
entries of the list & second estimator, one uniform entry per proposal, 30 seeds & $-74.43$ & $-74.48$ (99\%) \\
entries of the list & replica of~\cite{PR1809}, one uniform entry per batch, 64 runs & $-74.45$ & $-74.51$ (99.9\%) \\
real claws of the walk & the same replica on screened windows, $W_9\in\mathcal{F}_9$, $W_8\in\mathcal{F}_8$, 128 runs (\cref{sec:verification}) & $-74.45$ & $-74.49$ (99.9\%) \\
\bottomrule
\end{tabularx}
\end{table}

\subsection{The sign class of \texorpdfstring{$\sig{0}(W_{11})$}{sigma0(W11)}}\label{sec:class}

The word $W_{11}$ enters rows 16 to 38 only through its differences: the modular difference $D_{11} = \hex{0601dcd4}$ at $W_{18}$ and $W_{27}$, fixed by the characteristic, and the modular difference $\sig{0}(W_{11}') \boxminus \sig{0}(W_{11})$ at $W_{26}$, which is not. The published pair's $W_{11}$ has XOR difference \hex{06022334}, whose image under $\sig{0}$ has eight active positions, so the $W_{26}$ term takes one of 128 modular values, the \emph{class} of the pair, and exactly $2^{-7}$ of the words $W_{11}$ that carry $D_{11}$ fall in each class. The tails of 57 pairs found by SAT on the published characteristic fall into two clusters about one bit apart according to their class.

We measured every class with the second estimator, with the $W_{26}$ term of the partner set to each of the 128 values and everything else equal, on two seeds of $2^{27}$ proposals per class. Eight classes, the \emph{good} classes, measure $-74.37$ to $-74.44$; 32 classes, among them that of the published pair, measure $-75.25$ to $-75.46$; 16 measure about $-79.3$; and 72 give no collision. The tail is a function of the class alone in these measurements. The class is not forced by the characteristic: a complete enumeration of the dense part over the published core gives pairs in 12 classes, three of them good. Our own pair, measured in \cref{tab:tail}, is the first of class \hex{2000c1a0} in that enumeration, and the advice of the attack with four free words (\cref{tab:pair2}) is a pair of the good class \hex{e000be60}.

Each bit of the tail moves the time by one bit, so a good-class pair takes about one bit off the cost. The gain is specific to the carrier $K_P$: under the strict event of the characteristic, the advice of \cref{tab:pair2} and the published pair measure the same tail, and \cref{sec:gap} locates the gain. Nothing else in the cost changes with the pair: the claw supplies $W_8$, $W_9$ and $W_{16},\dots,W_{23}$ afresh, and $p_8$, $p_9$ and the row-17 event are unchanged.

\subsection{The priced event against the characteristic}\label{sec:gap}

The carrier accepts collisions that leave the characteristic. To measure how many, we extended the second estimator to the \emph{strict} event, which checks every row from 18 to 38 on its signed differences and on the conditions of the Boolean functions, with the two-bit relations of these conditions derived from the signed outputs of the pair in use. Four arms of 30 seeds each, with $2^{26}$ proposals per seed, were fixed with their seeds and their analysis before the runs, and every success is an SFS collision checked by two compressions. On the published pair the strict event measures $2^{-80.39}$ (one-sided 99\% bound $2^{-80.43}$), in agreement with the figure $2^{-80.41}$ of~\cite{GWZW26}. On the heads of the list attack, the words $W_8,\dots,W_{11}$ it produces, it measures $2^{-80.41}$ (bound $2^{-80.45}$), both under the law at which the tail is priced and on heads computed from uniform claws and entries; under $K_P$ the same heads give $2^{-74.43}$. The two events differ by $5.97$ bits on the heads of the list attack and by $5.05$ bits on the published pair, and under the strict event the good class gains nothing.

We located the difference by classifying every collision of the two relaxed arms, 135{,}054 on the advice and 72{,}224 on the published pair, against the strict conditions row by row, each weighted by its run's estimate divided by its run's number of collisions (\cref{tab:gap}). The classification reuses the registered runs and was not itself pre-registered; its totals lie within about $0.1$ bits of the direct measurements, the accuracy to which we read its factors.

\begin{table}[t]
\centering
\caption{The ratio of the relaxed to the strict event as three successive factors ($\log_2$, with 95\% intervals from resampling seeds). Each factor is the weighted fraction of relaxed collisions that meet the stated conditions among those that meet the conditions of the rows above.}\label{tab:gap}
\setlength{\tabcolsep}{5pt}
\begin{tabular}{@{}lrr@{}}
\toprule
Conditions & Advice & Published pair \\
\midrule
Signs of $\Sig{1}$ and $\IF$ at steps 20 and 21 & $3.75\ [3.66, 3.83]$ & $3.91\ [3.82, 4.02]$ \\
Signed differences at rows 24 to 38 & $1.16\ [1.12, 1.19]$ & $0.26\ [0.24, 0.28]$ \\
Other conditions at rows 24 to 38 & $0.99\ [0.96, 1.03]$ & $1.01\ [0.97, 1.06]$ \\
\midrule
Total & $5.89\ [5.79, 6.00]$ & $5.18\ [5.09, 5.29]$ \\
Direct measurements & $5.97$ & $5.05$ \\
\bottomrule
\end{tabular}
\end{table}

\paragraph{Steps 20 and 21.} On both pairs, 93\% of the relaxed collisions break at least one of seven conditions that fix the signed outputs of $\Sig{1}(E_{19})$ at step 20 and of $\Sig{1}(E_{20})$ and $\IF(E_{20},E_{19},E_{18})$ at step 21. The carrier checks the signed state differences at rows 19 to 23, so these deviations cancel in the modular sums of steps 20 to 22: the collisions follow other signed characteristics with the same state differences, and at these steps the priced event measures a differential rather than one characteristic.

\paragraph{After row 23.} A fraction $2^{-1.16}$ of the collisions that meet every condition through row 23 keeps the signed differences of the characteristic at rows 24 to 38 on the advice, against $2^{-0.26}$ on the published pair. The difference, $0.90$ bits, matches the advantage of the good class under $K_P$, while the other two factors differ by at most $0.16$ bits: the gain of the good class lies in collisions that leave the characteristic after row 23, where the class enters through $W_{26}$.

\section{The Collision Attack on 39-Step \sha{}}\label{sec:attack}

The attack outputs two 128-byte messages $M_0\,\|\,M_1$ and $M_0\,\|\,M_1'$ that share the first block and whose second blocks give the same chaining value; the padding block that follows is identical, so the digests are equal. Its only non-uniform advice is a 160-byte SFS pair on the characteristic of \cref{tab:char}. We present the attack with four free state words first (\cref{sec:alg,sec:listcost}) and state its premises in \cref{sec:listprem}. The attack with two free words, which the first valid filings implement, is the special case of a list that holds only the advice's entry; \cref{sec:two} gives its cost.

\subsection{The attack}\label{sec:alg}

\begin{algorithm}[h!]
\caption{Collision attack on 39-step \sha{} with four free state words}\label{alg:attack}
\begin{algorithmic}[1]
\State Expand the advice pair by two compressions; keep $A_4,\dots,A_7$, $E_8,\dots,E_{11}$ and $W_{12},\dots,W_{15}$ of both members, the differences $D_8,\dots,D_{12}$, $g$ and $w_{17}$.
\State Generate the list $L_\eta$ by rule as the 18-bit prefixes of its keys, grouped by $\kappa$ (\cref{sec:list}).
\For{$\lceil B/0.99308 \rceil$ windows}
\State Draw first blocks $M_0$ until one passes the screen; compute $\cv$ and the constants of $F$ (\cref{prop:claw}, with $c_9' \boxplus \eta$ for $c_9$).
\State Generate the bit-sliced program of $F$ for this $\cv$.
\State Run the two-stage distinguished-point search of \cref{sec:walk} for $2^{26}$ evaluations of $F$; stop the window at $W_{\max}$ operations.
\For{each distinct claw $(a_0,u)$ found with $W_9 \in \mathcal{F}_9$, up to 944}
\State Compute $a_1$ and $w$; run the pass of \cref{sec:list} over the prefixes.
\For{each key that satisfies \cref{eq:w17} and each of its mates with $W_8 \in \mathcal{F}_8$}
\State Build $M_1$ and $M_1'$, which differ by $D_8,\dots,D_{12}$ in words 8 to 12; compress both from $\cv$; \textbf{if} equal \textbf{return} $(M_0\|M_1, M_0\|M_1')$.
\EndFor
\EndFor
\EndFor
\end{algorithmic}
\end{algorithm}

\subsection{Complexity with four free words}\label{sec:listcost}

The figures are those of the reference implementation~\cite{PR1809}. Its program counts every operation in the record convention of \cref{sec:walk}, and we state each cost in compression-equivalents and in the board's model (\cref{sec:costmodels}), which differ by $3.77$ bits.

\paragraph{Trials.} A \emph{trial} is an entry with $W_8 \in \mathcal{F}_8$ in a claw with $W_9 \in \mathcal{F}_9$. A window of $2^{26}$ evaluations yields $57{,}003$ distinct claws on average, of which about 907 pass $\mathcal{F}_9$; at most 944 are evaluated per window. Each of these has on average $681{,}709$ entries in $\mathcal{F}_8$, about $9/128$ of the list. Priced at one-sided 99.5\% bounds on the two factors, $901.6$ claws and $681{,}192$ entries, a window holds $T = 2^{29.19}$ trials. The reference implementation counts entries on the first 125 claws of each window that pass $\mathcal{F}_9$; an exhaustive count on all such claws of 80 full windows moves the time by less than $0.001$ bits (\cref{app:impl}).

\paragraph{Probability per trial.} A trial yields a collision with probability $q\,\ptail$, where $q = 2^{-32}$. The reference implementation prices $\ptail = 2^{-74.46}$, the pooled mean of 14 organizer-run replicas of the estimator of \cref{sec:smc}; its one-sided 99.9\% bound is $2^{-74.66}$. The priced value is a mean and not a bound. Our own measurements give $2^{-74.43}$ and $2^{-74.45}$, with one-sided bounds of $2^{-74.48}$ at 99\% and $2^{-74.51}$ at 99.9\% (\cref{tab:tail}).

\paragraph{Success of a window.} Every key that satisfies \cref{eq:w17} is tested in every window, so a window succeeds on a union of tail events. The next proposition bounds its probability without any assumption on the dependence between tails.

\begin{proposition}[Success of a window]\label{prop:success}
Fix the claws of a window. Call a key a \emph{trial key} if it has a mate with $W_8\in\mathcal{F}_8$ in a claw with $W_9\in\mathcal{F}_9$. For a trial key $k$ let $X_k$ indicate that $k$ satisfies \cref{eq:w17}, let $U_k$ be the event that one of its mates with $W_8\in\mathcal{F}_8$ gives a collision, and let $p_k = \Pr[U_k]$. If $\Pr[X_k=1]=q$ for every trial key and the family $(X_k)$ is independent of the events $(U_k)$, then the window outputs a collision with probability at least
\[
q \sum_k p_k \Bigl(1 - \tfrac12\nu_k\Bigr), \qquad \nu_k = \mathbb{E}\Bigl[\,\sum_{l\ne k} X_l \Bigm| X_k = 1\Bigr].
\]
\end{proposition}

\begin{proof}
The program builds and compresses every mate with $W_8\in\mathcal{F}_8$ of every key that satisfies \cref{eq:w17}, so the window succeeds on $\bigcup_{k\in H}U_k$, where $H=\{k: X_k=1\}$. By the second Bonferroni inequality, and since $\Pr[U_k\cap U_l]\le\min(p_k,p_l)\le(p_k+p_l)/2$,
\[
\Pr\Bigl[\,\bigcup_{k\in H}U_k \Bigm| H\Bigr] \;\ge\; \sum_{k\in H}p_k - \sum_{\{k,l\}\subseteq H}\Pr[U_k\cap U_l] \;\ge\; \sum_{k\in H}p_k\Bigl(1-\frac{|H|-1}{2}\Bigr).
\]
The expectation of the right side over $H$ is $\sum_k p_k\,\mathbb{E}\bigl[X_k\bigl(1-\tfrac12\sum_{l\ne k}X_l\bigr)\bigr] = q\sum_k p_k\bigl(1-\tfrac12\nu_k\bigr)$.
\end{proof}

\paragraph{The terms of the bound.} Under premises H5 and H7 (\cref{sec:listprem}), keys of other claws contribute at most $0.121$ to $\nu_k$ and keys of the same claw at most $1.1\cdot 10^{-3}$, and further mates of a key lower $p_k$ below the sum over its mates by a factor of at most $1-10^{-4}$. With a margin of $10^{-4}$ for the selection of claws by the work cap, the first-order sum over the trials of a window is lowered by the factor $1 - 0.0616$, that is, by $0.09$ bits (\cref{app:proofs}).

\paragraph{Time.} Each window stops at a work cap of $W_{\max} = 573{,}465{,}000$ operations, asserted by the program, and a capped window keeps at least the fraction $0.99308$ of its trials, a one-sided 99\% bound on 2{,}160 predeclared windows that we recomputed from the per-window rows. Windows are independent, so after $B/0.99308$ windows the attack has failed with probability at most $\exp\bigl(-B\,T\,q\,\ptail\,(1-0.0616)\bigr)$. For success probability $0.39$ it needs
\[
B = \frac{-\ln 0.61 \cdot 2^{32+74.46}}{2^{29.19}\,(1-0.0616)} = 2^{76.34}
\]
windows, and its time is $(B/0.99308)\,(W_{\max}+6)$ operations, $2^{97.8}$ compression-equivalents or $2^{94.0}$ in the board's model. The one-time generation of the list ($2^{30}$ operations) and the construction of the advice (\cref{sec:advice}) do not change the rounded figures (\cref{tab:ledger94}).

\begin{table}[t]
\centering
\caption{The terms of the cost of the attack with four free words ($\log_2$).}\label{tab:ledger94}
\begin{tabular}{@{}lrr@{}}
\toprule
Term & Compression-equivalents & Board model \\
\midrule
$W_{17}=w_{17}$ per trial & $32.0$ & $32.0$ \\
Tail per trial & $74.5$ & $74.5$ \\
Trials per window & $-29.2$ & $-29.2$ \\
Work per window & $21.4$ & $17.6$ \\
Yield of the work cap & $0.0$ & $0.0$ \\
\midrule
Per expected collision & $98.7$ & $94.9$ \\
Dependence between tails (\cref{prop:success}) & $0.1$ & $0.1$ \\
\midrule
Expected cost of the first collision, at most & $98.8$ & $95.0$ \\
Success probability $0.39$ & $-1.0$ & $-1.0$ \\
\midrule
Time at success probability $0.39$ & $97.8$ & $94.0$ \\
\bottomrule
\end{tabular}
\end{table}

\paragraph{Per expected collision.} By linearity, the expected number of collisions in a window is at least $0.99308\,T\,q\,\ptail$, with no second-order term. The cost per expected collision is therefore $2^{98.7}$ compression-equivalents, $2^{94.9}$ in the board's model. Since windows are independent and each succeeds with probability at least $(1-0.0616)$ times its expected number of collisions, the expected cost of the first collision is at most $2^{98.8}$, for any dependence between the tails within a window. A generic birthday search priced in the same unit costs about $2^{128}$.

\paragraph{Memory.} The peak memory is $1{,}017{,}085$ bytes, below $2^{20}$. The list is not stored: the program holds the 62{,}400 prefixes as bit planes and rebuilds them by rule at start-up. The list accounts for $\log_2 9{,}701{,}592 = 23.2$ of the 32 bits of $W_{17}$; $8.8$ remain.

\paragraph{The first block.} Each first block's chaining value is scored before the kernel of $F$ is generated, at 187 operations per draw, and only blocks whose score predicts a short kernel are kept.

\subsection{The attack with two free words}\label{sec:two}

The attack with two free words takes the published pair as advice and uses no list: lines 7 to 10 of \cref{alg:attack} reduce to the tests $W_9 \in \mathcal{F}_9$, $W_8 \in \mathcal{F}_8$ and $W_{17} = w_{17}$ on each claw. A claw yields a collision with probability $p_8\, p_9\, 2^{-32}\, \ptail = 2^{-117.48}$, of which 9.81 bits are exact counts (\cref{lem:side}), 32 rest on the uniformity of $W_{17}$ over claws, and 75.67 on the measured tail of the published pair. At success probability $0.39$ the attack costs $2^{122.1}$ compression-equivalents, $2^{118.3}$ in the board's model, with a peak memory of $1{,}047{,}892$ bytes~\cite{PR574} (\cref{app:two}).

\subsection{Advice and preprocessing}\label{sec:advice}

In the literature's convention, a published pair is public and reusing it costs nothing. In the board's convention, every bit of advice must be constructed inside the accounting. The filings count the construction as follows: the six steps 10 to 15 carry 384 bits of state under 350 conditions of the characteristic, leaving $2^{34}$ candidates for the dense part, each tested for non-empty row-16 and row-17 fibres by exact bit-level enumeration in under $2^2$ compressions, so at most $2^{36}$ compressions in $2^{16}$ bytes. The corresponding figure in~\cite{LZL+26} is about $2^{34.2}$; that paper reports no cost for the search that produced the characteristic, and that search is not charged here. The list attack's advice was chosen from 93 such pairs by their tails; the reference implementation charges this choice, together with the search for the fibre, as $2^{43}$ compressions. Neither charge changes the rounded time.

\section{Comparison with Joint Message-Word Pinning}\label{sec:pinning}

Gong \emph{et al.}~\cite{GWZW26} attack 39-step \sha{} on the same characteristic with the published pair. They tabulate the preimages of $f(x) = \sig{0}(x) \boxminus x$, the odd branch $\Fodd$ of \cref{prop:claw}, in 16\,GiB with a 1\,GiB index. With $A_0$ and $A_3$ fixed, one look-up solves the row-16 equation of \cref{prop:claw} for $W_1$ at each of 300 values of $h$ (\cref{lem:four}), a second solves \cref{eq:w17} for $W_2$, and the solution is kept if its $A_2$ lies on the fibre. In the terms of \cref{sec:claw,sec:list}, look-ups take the place of the claw and $A_3$ is pinned, so that a solution of row 16 is tested against at most 4{,}096 values of $A_2$ rather than against a list of 9{,}701{,}592 entries. They report $2^{108.8}$ timed scalar compressions per expected collision ($2^{111.0}$ in single-thread SHA-NI units) with 17\,GiB of memory.

The two attacks price steps 18 to 38 on different events: Gong \emph{et al.} on the strict event of the characteristic, $2^{-80.41}$ on the published pair, and the list attack on the carrier $K_P$. We measured each event on the other attack's heads. The strict event gives $2^{-80.41}$ on the heads of the list attack (\cref{sec:gap}), and $K_P$ gives $2^{-75.33}$ (99\% bound $2^{-75.37}$) on heads drawn as joint pinning draws them. On the strict event the list attack therefore costs $2^{104.6}$ compression-equivalents per expected collision, where Gong \emph{et al.} report $2^{108.8}$ timed scalar compressions; on $K_P$ joint pinning costs $2^{103.7}$ timed scalar compressions, where the list attack costs $2^{98.7}$ compression-equivalents. The two units differ by the relative speed of the machines. For the first collision, Gong \emph{et al.} multiply the cost per expected collision by the largest number of solutions of one chaining value, $2^{15.84}$, which gives $2^{124.6}$ with no premise on how solutions cluster; \cref{prop:success} adds $0.09$ bits to the cost per expected collision, but rests on premises H5 and H7 (\cref{sec:listprem}).

\begin{table}[t]
\centering
\caption{Cost per expected collision of the two attacks timed on one machine ($\log_2$ of 39-step compressions; Apple M1 Pro, one core). Memory: $2^{20}$ bytes for the list attack, 17\,GiB for joint pinning. The last line is the figure of~\cite{GWZW26} on their machine.}\label{tab:timing}
\setlength{\tabcolsep}{5pt}
\begin{tabular}{@{}llrr@{}}
\toprule
Attack & Tail event & Scalar & ARM SHA \\
\midrule
List attack & carrier $K_P$ & $97.1$ & $100.4$ \\
Joint pinning, replayed & carrier $K_P$ & $102.4$--$102.6$ & $105.7$--$105.8$ \\
\addlinespace
List attack & strict & $103.0$ & $106.3$ \\
Joint pinning, replayed & strict & $107.5$--$107.6$ & $110.8$--$110.9$ \\
\midrule
Joint pinning, on their machine~\cite{GWZW26} & strict & $108.8$ & $111.0$ (SHA-NI) \\
\bottomrule
\end{tabular}
\end{table}

To compare in one unit, we timed both attacks on one core of an Apple M1 Pro (\cref{tab:timing}). For the list attack, a native port of a window generates the kernel of $F$ for each first block as the reference generator does and reproduces every count of the reference implementation; a window takes $104.9$\,ms. For joint pinning we replayed an entry from the description of~\cite{GWZW26}, with the dense part of the published pair recovered from their worked example and with their set of classes; an entry takes 11.3 to 12.5\,\textmu s, $1.2$ to $1.3$ bits less in scalar units than the entry timed in~\cite{GWZW26}. The unit is a 39-step compression with message schedule and feed-forward, $115.5$\,ns in scalar C and $11.85$\,ns with the ARMv8 SHA-256 instructions. On the strict event the list attack is 4.5 to 4.6 bits cheaper, or 4.3 bits if the look-ups of joint pinning are batched with explicit prefetch, and on $K_P$ it is 5.3 to 5.5 bits cheaper. Up to step 17, the list attack spends $2^{24.2}$ compression-equivalents per second block that meets every condition, against $2^{28.4}$ timed scalar compressions for joint pinning: the walk finds a solution of row 16 at a higher price than a look-up, and tests it against a set some 2{,}400 times larger. The replayed entry is a reimplementation, not their code, and its table is built at a quarter of the full size, which we estimate understates its cost by $0.03$ bits.

\section{Premises and Verification}\label{sec:verification}

The cost of \cref{sec:attack} is conditional on declared premises. This section states them, the experiments that test their components, and the quantities that remain untested; \cref{app:verif} gives the evidence in full.

\subsection{Premises}\label{sec:listprem}

The reference implementation declares the following premises~\cite{PR1809}; H1, H3 and H5 extend those of the attack with two free words. The costs stated in this paper use H1, H2, H3, H5 and H7.
\begin{description}
\item[H1, claw model.] The chaining value of a random first block that passes the screen is an independent random input to the walk and the later events, so windows are independent trials. The claws of a window that pass $\mathcal{F}_9$ have independent, identically distributed $(a_0,a_1)$, the trial count is the one measured on the predeclared windows, and $W_{17}$ is uniform over claws.
\item[H2, tail.] $\ptail \ge 2^{-74.46}$ for a uniform trial of the attack, the mean pooled over the classes of entries; the mean over entries of the advice's own class is $2^{-74.48}$. The value is a mean, measured on the estimator's own draws of $W_8$, $W_9$ and one uniform entry per batch (\cref{sec:tail}), and carried to the claws of the walk. For the attack with two free words the premise is $\ptail \ge 2^{-75.67}$ for claws uniform on the fibre.
\item[H3, advice.] The 160-byte pair can be constructed at the charged cost (\cref{sec:advice}).
\item[H5, hits.] Each key is hit with probability $2^{-32}$ in each claw, independently across claws and independently of the tails, and distinct claw groups, within and across windows, are independent trials.
\item[H7, mates.] At a hit, $W_3$ is uniform. Together with an exhaustive enumeration of rows 18 and 19, this bounds by $1.97\cdot 10^{-4}$ the probability that a second mate of the same key also follows the trail.
\end{description}
\Cref{prop:success} uses H5 and H7, the trial count uses H1, and the tail uses H2. A correlation between the walk's selection of claws and the tail would move the exponent one for one; it is measured below for claws of the walk that pass $\mathcal{F}_9$ and entries in $\mathcal{F}_8$, and selection by the event $W_{17}=w_{17}$ is beyond the scale of these measurements. Because $W_{17}$ shares $u = W_1$ with the row-16 equation, its uniformity over claws is the premise with the most direct route to a test.

The full attack has not been run, and no collision of the target has been found; every experiment tests a component of the attack.

\subsection{Experiments}\label{sec:experiments}

The board runs each filing's declared experiments in a bounded, networkless sandbox and strips the participants' observations from what its judge sees, so an experiment establishes only what its exit status and the organizer's recomputation establish. The reference implementation declares 16 experiments, all run by the organizer with every reported value hard-asserted: the transplant and the list on one entry of each of the 200 groups, four small windows in check mode with planted keys and a work cap, key hits at an event scaled to probability $2^{-7}$, and 14 tail replicas with every hit verified as an SFS collision. The attack with two free words is checked in the same way, and the published pair passes through its claw path and collides. \Cref{app:verif} lists these experiments.

Two of us ran four further measurements outside the sandbox, each pre-registered before its counted runs, with every run reported and every hit inverted to a chaining value and checked by two 39-step compressions: the tail on the estimator's draws (\cref{sec:tail}), the trial count on all claws (\cref{sec:listcost}), $W_{17}$ over claws, and the tail on real claws. The last tests premise H2 where the attack uses it. Each of 128 runs executes one window of the reference implementation on its attack path, with the first-block screen and the work cap, and takes $W_8,\dots,W_{11}$ from the claws that pass $\mathcal{F}_9$ and from uniform entries with $W_8\in\mathcal{F}_8$; each run shares the population of rows 18 to 23 with a run under the replica's own law. Real claws give $2^{-74.45}$ on 57{,}652 hits and the replica's law $2^{-74.44}$ on 58{,}025, a mean paired difference of $-0.009$ bits with the 95\% interval $[-0.025,\ +0.007]$, so the selection of claws by the walk and by $\mathcal{F}_9$ does not lower the tail measurably. The estimator pins $W_{17}$, so selection by a hit on $W_{17}$ is not sampled. On the 72{,}040 claws that pass $\mathcal{F}_9$ in 80 full windows, all 32 bits of $W_{17}$ are consistent with uniform. Running the attack up to step 17 on the saved claws of 108 windows gives 83 keys that meet $W_{17}=w_{17}$, against 78.3 expected, and \cref{app:tables} gives one of the resulting pairs from a real first block.

\subsection{Untested quantities}

At the attack's scale of $2^{76.3}$ windows, the following remain untested: the yield and cost of a window; the tail and $W_{17}$ under the walk's selection of claws, which are measured on 128 runs and on 80 windows; the tail under the selection by a hit on $W_{17}$, which no sandbox measurement samples; and the probability $2^{-32}$. The filings are fetchable at fixed commits, and the measurements made for this report are deposited with their programs and records (\cref{sec:repro}).

\section{Open Problems}\label{sec:discussion}

\begin{itemize}
\item \emph{The tail.} The tail of an entry is priced at a measured mean on uniform draws of the fibre; on real claws of the walk that pass the filters it is at most $0.025$ bits lower at 95\%, and the selection of claws by a hit on $W_{17}$ is beyond sandbox scale (\cref{sec:verification}). About 6 bits of the priced tail come from collisions that leave the characteristic (\cref{sec:gap}). Computing the probability of the differential through rows 19 to 23, summed over the signed paths that give the same state differences, would replace part of the measured premise by a counted one. The dependence of the tail on the class of $\sig{0}(W_{11})$ is located but not derived from the $W_{26}$ and $W_{27}$ relations.
\item \emph{Timing and memory.} Both attacks are timed on one core of one machine (\cref{sec:pinning}); multi-core scaling and an x86 machine with SHA-NI remain unmeasured. Once the one-MiB constraint of the board is dropped, replacing the walk with two look-ups per claw in a table of the odd branch of $F$ would lower the cost of a window by an estimated 0.9 to 2.0 bits on that machine, depending on whether the look-ups are batched, at 8.5 to 17\,GiB of memory; this variant has not been implemented.
\item \emph{The rest of $W_{17}$.} The list of \cref{sec:list} covers 23.2 of the 32 bits. A later filing~\cite{PR1819} re-bases it to $\eta = \hex{3cc9e0be}$, on which the rule of \cref{sec:list} returns 12{,}025{,}320 entries on 936 values of $a_3$; our program gives the same count. On this fibre the rule is incomplete: our enumeration over all $a_3$ gives 12{,}574{,}440 entries on 1{,}872 values, or 23.58 bits, because the rule leaves out the 936 values of $a_3$ that put $E_7$ outside its 14 bits, and each of their 549{,}120 entries keeps both members on the advice's states through step 15. The tail has not been measured on the entries of this fibre. With this fibre and further engineering, the board's best stood at $2^{93.62}$ in the board's model on 21 November~\cite{PR1838}, priced under premise H-cross-claw and with the claws per window and the yield of the work cap at measured points rather than bounds. Which fibre gives the largest list, and whether a further free state word covers more of $W_{17}$, is open.
\item \emph{The search on SHA-512.} The estimate of $2^{101}$ for 39-step SHA-512 (\cref{sec:sha512}) has no bound on the first collision under the clustering of hits, and the tail on real first blocks has not been compared with the sampled one there. The search does not yet use mates: classes with the same $\kappa$ share one value of $W_{10}$ per $a_2$, so one search over their union could serve several classes. A meet-in-the-middle split of the bits of $a_2$ was not tried. Whether the search pays at 38 steps, where step 17 costs $2^{-19.0}$~\cite{GWZW26}, is open.
\end{itemize}

\section{Conclusion}\label{sec:conclusion}

The 39-step characteristic of Li \emph{et al.}~\cite{LZL+26} was judged ineffective for \sha{} because every condition from step 16 onward had to be satisfied by trial. Leaving the first two state words free turns the 32-bit condition on $W_{16}$ into a claw that a distinguished-point search finds in about $2^{11}$ evaluations, and the transplant onto a random first block is a bijection, so the price of that freedom is exact. Leaving the next two state words free on a fibre of the function through which they reach row 16 turns the trial on $W_{17}$ into a test against a list of 9.7 million second blocks. The resulting attack costs about $2^{98.7}$ compression-equivalents per expected collision with one MiB of memory, and the expected cost of its first collision is at most $2^{98.8}$ for any dependence between the tails of its trials. Timed on one machine, it is 4.5 to 5.5 bits cheaper on either tail event than joint message-word pinning~\cite{GWZW26}, which needs a 17\,GiB table. The cost rests on declared heuristic premises, the most consequential of which is the measured probability of the uncontrolled steps: our measurements bound it from below by $2^{-74.51}$ with 99.9\% confidence on uniform draws, and on real claws of the walk that pass the filters it is at most $0.025$ bits lower at 95\% confidence (\cref{sec:verification}). On 39-step SHA-512, where row 17 admits $2^{30.4}$ values of $W_{17}$, searching the list bit by bit gains 29 bits over single-word pinning, for an estimated $2^{101}$ compressions per expected collision; this estimate also rests on a measured tail, and no collision has been found.

\section*{Acknowledgements}
We thank Eigen Labs and Shielded Labs, the organizers of the HashSmash board; Sebastian R. Wain (sraiw), whose filing 793185f6~\cite{PR421} is the construction on which the attack with two free words is built; and lamb356, whose filing with dariolina~\cite{PR574} is the reference implementation of that attack (\cref{app:provenance}). The filings were written with AI coding agents, as declared in their public notes (\cref{app:provenance}), and this report was drafted with AI assistance; the authors take full responsibility for every claim.

\bibliographystyle{splncs04}
\bibliography{refs}

\appendix
\renewcommand{\theHsection}{appendix.\Alph{section}}
\clearpage
\begin{center}\large\bfseries Supplementary Material\end{center}
\noindent The appendices are supplementary material. The paper can be read without them.
\section{Proofs}\label[appendix]{app:proofs}

\begin{proof}[of \cref{lem:transplant}]
Each step is invertible given its output word (\cref{sec:prelim}). At step 0, $T_0 = a_0 \boxminus \tau_0$, so $E_0 = A_{-4} \boxplus T_0$ and $W_0 = T_0 \boxminus E_{-4} \boxminus \Sig{1}(E_{-1}) \boxminus \IF(E_{-1},E_{-2},E_{-3}) \boxminus K_0 = a_0 \boxplus k_0$. At step 1, $T_1 = a_1 \boxminus \Sig{0}(a_0) \boxminus \MAJ(a_0,A_{-1},A_{-2})$ and $W_1 = T_1 \boxminus E_{-3} \boxminus \Sig{1}(E_0) \boxminus \IF(E_0,E_{-1},E_{-2}) \boxminus K_1$, which is the stated relation between $a_1$ and $u$. Steps 2 to 9 produce the pinned words $A_2,\dots,A_5$ and $E_6,\dots,E_9$ and determine $W_2,\dots,W_9$; at step 9, $E_5 = a_1 \boxplus A_5 \boxminus \Sig{0}(A_4) \boxminus \MAJ(A_4,A_3,A_2)$ is the only input that depends on the free words, which gives $W_9 = c_9 \boxminus a_1$. From step 10 onward, the state and the message words are the published ones.
\end{proof}

\begin{proof}[of \cref{thm:bijection}]
Rearranging the step function (\cref{sec:prelim}), for $i \ge 0$,
\begin{align*}
E_i &= A_i \boxplus A_{i-4} \boxminus \Sig{0}(A_{i-1}) \boxminus \MAJ(A_{i-1},A_{i-2},A_{i-3}),\\
W_i &= E_i \boxminus A_{i-4} \boxminus E_{i-4} \boxminus \Sig{1}(E_{i-1}) \boxminus \IF(E_{i-1},E_{i-2},E_{i-3}) \boxminus K_i .
\end{align*}
The words $A_2,\dots,A_5$ and $E_6,\dots,E_9$ are pinned.

\emph{Stages 1 and 2.} The word $E_5 = a_1 \boxplus A_5 \boxminus \Sig{0}(A_4) \boxminus \MAJ(A_4,A_3,A_2)$ depends on $a_1$ only, with unit slope, and every other term of $W_9$ is pinned, so $W_9 = c_9 \boxminus a_1$. With $a_1$ fixed, the word $E_4 = a_0 \boxplus A_4 \boxminus \Sig{0}(A_3) \boxminus \MAJ(A_3,A_2,a_1)$ has unit slope in $a_0$, and every other term of $W_8$ is pinned or a function of $a_1$. Neither word involves $\cv$. Hence $(a_0,a_1) \mapsto (W_8,W_9)$ is a bijection of $(\mathbb{Z}/2^{32})^2$.

\emph{Stage 3.} Fix $(a_0,a_1)$; then $E_4,\dots,E_7$ are fixed. In
\begin{align*}
E_3 &= A_3 \boxplus A_{-1} \boxminus \Sig{0}(A_2) \boxminus \MAJ(A_2,a_1,a_0),\\
E_2 &= A_2 \boxplus A_{-2} \boxminus \Sig{0}(a_1) \boxminus \MAJ(a_1,a_0,A_{-1}),\\
E_1 &= a_1 \boxplus A_{-3} \boxminus \Sig{0}(a_0) \boxminus \MAJ(a_0,A_{-1},A_{-2}),\\
E_0 &= a_0 \boxplus A_{-4} \boxminus \Sig{0}(A_{-1}) \boxminus \MAJ(A_{-1},A_{-2},A_{-3}),
\end{align*}
$E_{3-j}$ has unit slope in $A_{-1-j}$ and otherwise involves only $A_{-1},\dots,A_{-j}$, so $(A_{-1},\dots,A_{-4}) \mapsto (E_3,\dots,E_0)$ is a bijection. For $i = 7,6,5,4$ the word $W_i$ has slope $-1$ in $E_{i-4}$, and its other variable terms, $\Sig{1}(E_{i-1})$ and $\IF(E_{i-1},E_{i-2},E_{i-3})$, involve only $E_3,\dots,E_{i-3}$ besides fixed words. So $(E_3,\dots,E_0) \mapsto (W_7,\dots,W_4)$ is a bijection, solved one word at a time in this order.

\emph{Stage 4.} Fix also $A_{-1},\dots,A_{-4}$, hence $E_0,\dots,E_3$. For $i = 3,2,1,0$ the word $W_i$ has slope $-1$ in $E_{i-4}$ and otherwise involves only $E_{-1},\dots,E_{i-3}$ besides fixed words, so $(E_{-1},\dots,E_{-4}) \mapsto (W_3,\dots,W_0)$ is a bijection. With stage 3, for every $(a_0,a_1)$ the map $\cv \mapsto (W_0,\dots,W_7)$ is a bijection of $(\mathbb{Z}/2^{32})^8$.

\emph{Stage 5.} Fix $W_8,\dots,W_{15}$. Given $(W_{16},\dots,W_{23})$, the expansion read downwards,
\[
W_{r-16} = W_r \boxminus \sig{0}(W_{r-15}) \boxminus W_{r-7} \boxminus \sig{1}(W_{r-2}), \qquad r = 23, 22, \dots, 16,
\]
returns $W_7, W_6, \dots, W_0$ in turn: $W_{r-15}$ is $W_8$ for $r=23$ and the word just obtained for $r<23$, $W_{r-7}$ is among $W_9,\dots,W_{16}$, and $W_{r-2}$ is among $W_{14},\dots,W_{21}$. So $(W_0,\dots,W_7) \mapsto (W_{16},\dots,W_{23})$ is a bijection.

Hence $\Phi$ is the bijection $(\cv,a_0,a_1) \mapsto (W_8,W_9,W_0,\dots,W_7)$ of stages 1 to 4 followed, for each $(W_8,W_9)$ and the pinned $W_{10},\dots,W_{15}$, by the bijection of stage 5 on the last eight coordinates.
\end{proof}

\begin{proof}[of \cref{lem:side}]
The map $\sig{0}$ is linear over $\mathrm{GF}(2)$, and $\sig{0}(w)[i] = w[i+7] \oplus w[i+18] \oplus w[i+3]$, with the first two indices taken modulo 32 and the third term absent for $i \ge 29$. If $w'$ and $w$ differ exactly in a set $S$ of bit positions, then $\sig{0}(w')$ and $\sig{0}(w)$ differ exactly in the positions $\Delta$ of $\sig{0}\bigl(\sum_{j\in S}2^j\bigr)$, and $\sig{0}(w') \boxminus \sig{0}(w) = \sum_{i\in\Delta}\varepsilon_i 2^i$ with $\varepsilon_i = +1$ if $\sig{0}(w)[i]=0$ and $-1$ otherwise.

\emph{$\mathcal{F}_8$.} $D_8 = 2^{30}$ and $\hex{087ff000} = 2^{27}+2^{23}-2^{12}$. If $w[30]=0$ then $S=\{30\}$ and $\Delta = \{12,23,27\}$; the signs $(+,+,-)$ on $(27,23,12)$ are the only ones with this sum, so $w \in \mathcal{F}_8$ if and only if
\[
w[2]\oplus w[13] = 0, \qquad w[9]\oplus w[26] = 0, \qquad w[15]\oplus w[19] = 1.
\]
If $w[30]=1$, the addition carries into bit 31 and stops there, so $S=\{30,31\}$ and $\Delta=\{12,13,23,24,27,28\}$. Two adjacent positions $i, i+1$ contribute $\pm 2^i$ or $\pm 3\cdot 2^i$, so the sum is reached only with the signs $(+,-)$ on $(28,27)$, $(+,-)$ on $(24,23)$ and $(-,+)$ on $(13,12)$. This gives the three conditions above together with
\[
w[3]\oplus w[14]\oplus w[31] = 0, \quad w[10]\oplus w[27]\oplus w[31] = 0, \quad w[16]\oplus w[20]\oplus w[31] = 1.
\]
In each case the conditions are independent, so $p_8 = 2^{-4} + 2^{-7} = 9/128$. The 14 bits involved are those of \hex{CC19E60C}, and $2^{14}\cdot 9/128 = 1152$.

\emph{$\mathcal{F}_9$.} $D_9 = 2^{12}-2^{23}-2^{27}$ and $\hex{03010000} = 2^{26}-2^{24}+2^{16}$. If $w[12]=0$ and $w[23]=w[27]=1$, no carry occurs, $S=\{12,23,27\}$, and $\Delta=\{16,24,26\}$, the positions 5, 9 and 20 cancelling in pairs. The signs $(+,-,+)$ on $(26,24,16)$ are the only ones with this sum, which gives
\[
w[1]\oplus w[12]\oplus w[29] = 0, \quad w[10]\oplus w[27]\oplus w[31] = 1, \quad w[2]\oplus w[19]\oplus w[23] = 0,
\]
six independent conditions in all, satisfied by $2^{-6}$ of the words. The other carry patterns are settled by enumeration: of the $2^{17}$ assignments of the bits selected by \hex{F9983C0F}, 2080 pass, the 2048 of the pattern without carry and 32 more, and membership does not depend on the other 15 bits. So $p_9 = 2080/2^{17} = 65/4096$.

By stages 1 and 2 of \cref{thm:bijection}, $(W_8,W_9)$ is uniform over uniform $(a_0,a_1)$ for every $\cv$, so the two events are independent.
\end{proof}

The conditions of this proof agree with $\mathcal{F}_8$ on $2^{20}$ random words, with both counts on all assignments of the active bits, and with an exhaustive enumeration of both sets over all $2^{32}$ words.

\begin{proof}[of \cref{lem:four}]
We use the two relations at the start of the proof of \cref{thm:bijection}. The pinned words determine $A_8,\dots,A_{11}$, so after step 11 the state is the advice's, and $W_{12},\dots,W_{15}$ are the advice's. The first relation gives $E_7 = a_3 \boxplus c_7$, $E_6 = a_2 \boxplus t(a_3)$, $E_5 = a_1 \boxplus A_5 \boxminus \Sig{0}(A_4) \boxminus \MAJ(A_4,a_3,a_2)$ and $E_4 = a_0 \boxplus A_4 \boxminus \Sig{0}(a_3) \boxminus \MAJ(a_3,a_2,a_1)$, while $E_0$ and $E_1$ are functions of $(\cv,a_0,a_1)$ and $E_2 = a_2 \boxplus A_{-2} \boxminus \Sig{0}(a_1) \boxminus \MAJ(a_1,a_0,A_{-1})$.
\begin{enumerate}
\item $W_0$ and $W_1$ involve only $E_0$, $E_1$, the chaining value and $a_0$, $a_1$.
\item $W_2 = E_2 \boxminus A_{-2} \boxminus E_{-2} \boxminus \Sig{1}(E_1) \boxminus \IF(E_1,E_0,E_{-1}) \boxminus K_2$, in which only $E_2$ involves $a_2$, with unit slope, and nothing involves $a_3$.
\item $W_9 = E_9 \boxminus A_5 \boxminus E_5 \boxminus \Sig{1}(E_8) \boxminus \IF(E_8,E_7,E_6) \boxminus K_9$; substituting $E_5$ gives the stated form.
\item $W_{10} = E_{10} \boxminus A_6 \boxminus E_6 \boxminus \Sig{1}(E_9) \boxminus \IF(E_9,E_8,E_7) \boxminus K_{10}$; substituting $E_6$ gives the stated form.
\item $W_{11} = E_{11} \boxminus A_7 \boxminus E_7 \boxminus \Sig{1}(E_{10}) \boxminus \IF(E_{10},E_9,E_8) \boxminus K_{11}$ involves the free words through $E_7$ only, and $W_8 = E_8 \boxminus A_4 \boxminus E_4 \boxminus \Sig{1}(E_7) \boxminus \IF(E_7,E_6,E_5) \boxminus K_8$ involves $E_4,\dots,E_7$, none of which involves $\cv$.
\end{enumerate}
\end{proof}

We also checked the identities on 20{,}000 random $(\cv,a_0,a_1,a_2,a_3)$ with a program independent of the filings; at the advice's own words, the transplant returns the advice block.

\begin{proof}[of \cref{prop:smc}]
For a state $x$ through row $t-1$ and a function $f$ of states through row $t$, let $(Q_t f)(x) = \mathbb{E}\bigl[\mathbf{1}_{R_t}\, f(x, E_t)\bigr]$ over a uniform $E_t$, where $R_t$ is the event that row $t$ of $K_P$ and the look-ahead hold. A candidate drawn with its $n_t$ fixed bits set is uniform on the $2^{-n_t}$ fraction of values that agree with them, outside which $R_t$ fails, so $\mathbb{E}\bigl[2^{-n_t}\mathbf{1}_{\text{accepted}}\, f\bigr] = (Q_t f)(x)$ for each candidate of a particle in state $x$. Write $\eta_t(f)$ for the mean of $f$ over the population after stage $t$, and condition on that population after stage $t-1$. Because the new population is drawn uniformly among the accepted candidates,
\begin{align*}
\mathbb{E}\bigl[\rho_t\,\eta_t(f)\bigr] &= \mathbb{E}\Bigl[\frac{2^{-n_t} a_t}{N m_t}\cdot\frac{1}{a_t}\sum_{c\ \text{accepted}} f(c)\Bigr]\\
&= \frac{1}{N}\sum_{i=1}^{N} (Q_t f)(x_i) = \eta_{t-1}(Q_t f),
\end{align*}
which also holds when $a_t = 0$. Applying this identity for $t = 23, 22, \dots, 18$ and taking expectations in turn gives
\[
\mathbb{E}\bigl[\rho_{18}\cdots\rho_{23}\,\eta_{23}(f)\bigr] = (Q_{18}\cdots Q_{23} f)(x_{17}),
\]
where $x_{17}$ is the pinned state. Let $g(x)$ be the probability, over $W_8$ uniform on $\mathcal{F}_8$ and $W_9$ on $\mathcal{F}_9$, that a state $x$ through row 23 ends in a collision. Since every particle drives the same number of proposals, $\mathbb{E}[h/P] = \eta_{23}(g)$ given the population, so $\mathbb{E}[\widehat{Q}] = (Q_{18}\cdots Q_{23}\, g)(x_{17})$. The look-ahead removes only states through row $t$ that fail row $t+1$, or the zero difference of $W_{t+2}$, for every value of $E_{t+1}$, because the modular differences it tests do not depend on $E_{t+1}$; so the composition equals $\Pr_U[\text{collision}\wedge K_P]$. This is the unbiasedness of the unnormalized Feynman--Kac particle estimator~\cite{DelMoral04}.
\end{proof}

\paragraph{The terms of the bound of \cref{prop:success}.} The quantity $\nu_k$ has two parts. Keys of other claws are hit independently of $k$ (premise H5 of \cref{sec:listprem}); they contribute at most $(s-1)\,\mu_D\,q$ for $s$ evaluated claws with $\mu_D$ trial keys each on average; with the measured $s-1 \le 898.1$ and $\mu_D \le 579{,}271$ this is $0.121$. Two keys of the same claw are hit together exactly when their values of $\Fodd(a_2 \boxplus w) \boxplus \kappa$ coincide; the mean number of other keys of the same claw with which a key coincides, measured on 16{,}384 values of $w$, is at most $1.1\cdot 10^{-3}$, and keys that coincide with four or more others are left out of the sum at a cost of at most $3\cdot 10^{-4}$ of it. Within a key, the same inequality applied to its mates gives $p_k \ge (1-\varepsilon/2)\sum_j \Pr[T_j]$, where $T_j$ is the tail event of mate $j$ and $\varepsilon \le 2\cdot 10^{-4}$ bounds the expected number of further mates that follow the trail given one that does (premise H7). With a margin of $10^{-4}$ for the selection of claws by the work cap, the first-order sum $q\sum_j \Pr[T_j]$ over the trials of a window is lowered by the factor $1 - 0.0616$, that is, by $0.09$ bits.

\section{Implementation and Measurements}\label[appendix]{app:impl}

\subsection{The walk}\label{app:walk}

The walk of \cref{sec:walk} runs in two stages.

\paragraph{Stage 1.} The walk advances 256 trails in parallel, one evaluation of $F$ each per iteration, for $2^{18}$ iterations. A point is distinguished when its low 7 bits are zero. Its 25 high bits select a slot in a table of 26{,}112 words of 32 bytes holding six 5-byte entries each (156{,}672 slots); an entry stores 8 check bits, a 12-bit trail length and a 20-bit trail identifier, and the slot and the check determine the point, so there are no false matches. A matching entry means that two trails have reached the same distinguished point; the pair is queued (up to 4096 pairs) and the lane starts a new trail. Trails older than 2048 steps are cut by a scan every 1024 iterations.

\paragraph{Stage 2.} When 3840 pairs are queued, and at the end of the window, the queued pairs are re-walked in 128 slots, the longer trail first and the shorter joining once the length difference has been covered. The first equal values identify the two predecessors, which form a claw if their parities differ. With two free words, each claw is tested against the filters of \cref{lem:side} and the condition $W_{17}=w_{17}$, and a claw that passes all three is built into a pair of second blocks, which are compressed and compared; with four free words, each claw that passes $\mathcal{F}_9$ is tested against the list (\cref{sec:list}).

\paragraph{Yield.} With two free words, over 24 full windows on the real step function the walk produced 1{,}213{,}630 distinct claws, 50{,}567.9 per window (standard deviation 186.7). Per distinct claw it took 1327.1 stage-1 and 893.8 stage-2 evaluations of $F$, in total $2221 \approx 2^{11.1}$ evaluations, with 10.37 distinguished-point hits, 2.540 matches, 1.250 opposite-parity claws and 1.290 same-parity collisions. The measured pass rates were 24{,}069 of 1{,}517{,}394 raw claws for $\mathcal{F}_9$ ($0.0159$ against $65/4096 = 0.0159$) and 1{,}735 of 24{,}069 for $\mathcal{F}_8$ ($0.0721$ against $9/128 = 0.0703$); no claw passed the test on $W_{17}$, as expected at probability $2^{-32}$.

\paragraph{Implementation.} The walk evaluates $F$ for 256 lanes at once as a straight-line program of 5-byte records, each an XOR, AND, OR or NOT of two 256-bit registers into a third. Bit $j$ of the values of all 256 lanes is held in register plane $j$, so rotations amount to renaming and cost nothing. The constants of $F$ are folded in as constant bits, so the program is generated for each first block, with every operation of the generator charged. The kernel has 690 to 751 records, and one evaluation of $F$ costs 2.8 to 3.7 operations per lane, compared with 205.6 for a compression. Every operation runs through one counted machine class, so the cost is read from the program's counter: 10{,}263 operations per distinct claw in the reference implementation with two free words (3.65 compressions in the board's model, 49.9 compression-equivalents), and about 8{,}320 with cheaper bookkeeping (2.96 and 40.5). The program counts every byte it holds; the distinguished-point table is the largest part, and the peak stays under $2^{20}$ bytes.

\paragraph{The walk with four free words.} The attack of \cref{sec:attack} runs the same walk with a table of 99{,}328 slots, stage 2 started at 3{,}328 queued pairs, trails cut at 1{,}536 steps, and the work cap of \cref{sec:listcost}. On three windows of its reference implementation, replayed with event counters (\cref{sec:pinning}), the walk yields 56{,}492 to 56{,}958 distinct claws, against 57{,}003 on average in \cref{sec:listcost}, at 1{,}882 to 1{,}918 evaluations of $F$ per distinct claw, about 1{,}190 in stage 1 and 710 in stage 2. On the first of them it costs 8{,}217 operations per distinct claw, or 40.0 compression-equivalents.

The reference implementation with two free words checks $W_{16}=g$ on every raw claw, with no failure on 1{,}517{,}394 claws.

\subsection{The list}

The reference implementation reports that the rule of \cref{sec:list} returns the same set as an enumeration over all $2^{32}$ values of $a_3$ with the trail of both members checked on every entry. We rebuilt the rule with an independent program and obtained the same counts. We also repeated the enumeration. Up to step 15, $a_2$ and $a_3$ affect the differences only through $\IF$ at steps 9 and 10, so for each of the $2^{32}$ values of $a_3$ we kept the pairs on the fibre for which those two steps give the advice's states in the second member and $W_{10}\in\mathcal{F}_{10}$, $W_{11}\in\mathcal{F}_{11}$, and ran every pair kept through step 15 on both members, from a random chaining value and random $(a_0,a_1)$. Each has the advice's states from step 8 on and no difference before it, and the set is the rule's. The agreement depends on the fibre (\cref{sec:discussion}).

\subsection{The trial count}

The reference implementation counts entries on the first 125 claws of each window that pass $\mathcal{F}_9$. We counted them exhaustively on all such claws of 80 full windows, $72{,}040$ claws in all: a claw has $681{,}671$ entries in $\mathcal{F}_8$ on average, with the 95\% interval $[681{,}205,\ 682{,}137]$ over the window means and a standard deviation of $10.3\%$ per claw, and the first 125 claws of a window do not differ measurably from the rest. Priced on all claws, the time rises by less than $0.001$ bits.

\subsection{The attack with two free words}\label{app:two}

The attack with two free words takes the published pair as advice and uses no list: lines 7 to 10 of \cref{alg:attack} reduce to the tests $W_9 \in \mathcal{F}_9$, $W_8 \in \mathcal{F}_8$ and $W_{17} = w_{17}$ on each claw, with the words computed by \cref{lem:transplant}. A claw yields a collision with probability
\[
\pclaw = p_8\, p_9\, 2^{-32}\, \ptail, \qquad \log_2 \pclaw = -9.81 - 32 - 75.67 = -117.48.
\]
Of these 117.48 bits, 9.81 are exact counts (\cref{lem:side}), 32 rest on the uniformity of $W_{17}$ over claws, and 75.67 on the measured tail of the published pair. Two claws of one window can depend on each other only if they share an endpoint, so claws are grouped by shared endpoints and one claw per group is counted (1{,}211{,}719 groups in 1{,}213{,}630 claws over 24 windows). Success probability $0.39$ then requires $2^{116.46}$ distinct claws, that is $2^{100.84}$ windows at the one-sided 99\% lower bound of 50{,}472.7 distinct claws per window. With the total work of the windows bounded by Hoeffding's inequality, the time is $2^{122.1}$ compression-equivalents, $2^{118.3}$ in the board's model, with a peak memory of $1{,}047{,}892$ bytes~\cite{PR574}. A tail measured on a larger replica ($2^{-75.47}$) and cheaper bookkeeping of the walk give $2^{121.6}$, and a pair of a good class (\cref{sec:class}) $2^{120.5}$.

\subsection{The price of free words on the published characteristics}\label{app:price}

The characteristics and SFS pairs of \cref{tab:criterion} are Tables 3 to 26 of~\cite{LZL+26}, Tables 8 to 11 of~\cite{ZMLM26} and Tables 12 and 14 of~\cite{GWZW26}. Each pair collides at its number of steps and gives its printed hash. The 39-step SHA-512 pair does so with $M_{14}$ and $M_{15}$ exchanged in both messages, as~\cite[App.~C.4]{GWZW26} notes, and with the leading zero of $M_3$ restored from $M'$.

For $m$ free words, the program pins the first member's words $A_m$ to $A_{m+3}$, $E_{m+4}$ to $E_{m+7}$ and $W_{m+8}$ to $W_{15}$ to the pair. It draws the chaining value and $A_0$ to $A_{m-1}$ uniformly, and gives the second member the pair's modular message difference. A draw counts when the second member reaches the pair's state after step $m+7$ and the $\sig{0}$ differences of $W_0,\dots,W_{m+7}$ are the pair's; every other word of the expansion carries the free words only linearly. Each run first checks the pair's own words. With $m=2$ the program gives $2^{-9.808}$ on the 39-step SHA-256 pair, against $585/2^{19} = 2^{-9.808}$ (\cref{lem:side}), and $2^{-9.00}$ and $2^{-13.02}$ on the 39- and 38-step SHA-512 pairs, which match the head factors measured in~\cite{GWZW26}.

\Cref{tab:price} splits the price into the state part and the $\sig{0}$ part. For the list's words the chaining value, $A_0$ and $A_1$ are held at the pair's values, only $A_2$ and $A_3$ are drawn, and the $\sig{0}$ conditions on $W_8$ and $W_9$ are left out, since the attack tests them per claw and per entry. The checks of \cref{prop:claw,lem:four} on random pinned states, and of the separability of $W_{17}$, use the same functions.

\begin{table}[t]
\centering
\caption{The price of free words in parts ($-\log_2$, $2^{28}$ draws; for the list's words on the 39-step SHA-256 pair, $2^{33}$ draws and 33 hits). A dash: no hit, so above 26.6. The joint price of $A_2$, $A_3$ on 39-step SHA-512 is measured with the rule of the list: the state part is exactly $2^{-21}$, and given it the $\sig{0}$ part holds in 130{,}890 of $2^{30}$ draws ($2^{-13.0}$); the unconditional $\sig{0}$ column overstates it because the two parts are correlated.}\label{tab:price}
\setlength{\tabcolsep}{4pt}
\small
\begin{tabular}{@{}llrrrrrr@{}}
\toprule
 & & \multicolumn{3}{c}{$A_0$, $A_1$} & \multicolumn{3}{c}{$A_2$, $A_3$} \\
\cmidrule(lr){3-5}\cmidrule(l){6-8}
Hash & Characteristic & state & $\sig{0}$ & joint & state & $\sig{0}$ & joint \\
\midrule
SHA-256 & 39~\cite{LZL+26} & $0.00$ & $9.81$ & $9.81$ & $8.00$ & $19.82$ & $27.96$ \\
SHA-512 & 39~\cite{LZL+26} & $0.00$ & $9.00$ & $9.00$ & $21.13$ & $20.86$ & $34.0$ \\
SHA-256 & 38~\cite{LZL+26} & $3.00$ & $13.77$ & $19.04$ & -- & -- & -- \\
SHA-256 & 38~\cite{GWZW26} & $4.00$ & $13.79$ & $17.60$ & -- & -- & -- \\
SHA-512 & 38~\cite{LZL+26} & $1.00$ & $12.01$ & $13.02$ & -- & -- & -- \\
SHA-256 & 37~\cite{LZL+26} & $13.00$ & -- & -- & -- & $21.57$ & -- \\
SHA-256 & 37~\cite{ZMLM26} & -- & -- & -- & -- & -- & -- \\
SHA-512 & 37~\cite{LZL+26} & $21.23$ & -- & -- & -- & $20.17$ & -- \\
SHA-256 & 36~\cite{LZL+26} & -- & -- & -- & -- & $13.78$ & -- \\
SHA-256 & 36~\cite{ZMLM26} & -- & -- & -- & -- & -- & -- \\
SHA-512 & 36~\cite{LZL+26} & -- & -- & -- & -- & $14.01$ & -- \\
\bottomrule
\end{tabular}
\end{table}

\subsection{The carrier by row}

\Cref{tab:carrierrows} gives the rate of each row of the carrier $K_P$ on entries of the list, from the 64 pre-registered runs of \cref{sec:tail}.

\begin{table}[t]
\centering
\caption{Rates of the carrier $K_P$ by row on entries of the list ($-\log_2$, mean of the per-run rates over 64 runs). Rows 18 to 23 are the factors of the sequential Monte Carlo, including the look-ahead; the later rows are the fractions of proposals whose state differences still equal the advice's. Rows 24, 27 and 29 to 38 pass with probability 1 once the earlier rows pass.}\label{tab:carrierrows}
\setlength{\tabcolsep}{4pt}
\small
\begin{tabular}{@{}lrrrrrrrrrr@{}}
\toprule
Row & 18 & 19 & 20 & 21 & 22 & 23 & 25 & 26 & 28 & Total \\
\midrule
$K_P$ & $18.42$ & $14.00$ & $15.20$ & $5.00$ & $8.00$ & $1.00$ & $1.00$ & $8.00$ & $3.84$ & $74.45$ \\
\bottomrule
\end{tabular}
\end{table}

\subsection{The list on 39-step SHA-512}\label{app:sha512}

The measurements of \cref{sec:sha512} use the pair, characteristic and extra conditions of~\cite[Tables~24--26]{LZL+26} with the corrections of~\cite[App.~C.4]{GWZW26}, and the pinning target and step-17 count of~\cite[Sect.~5.3]{GWZW26}. The first member is the message printed as $M'$. With $M_{14}$ and $M_{15}$ exchanged (\cref{app:price}), the pair meets every row of the characteristic. Three printed relations of the extra conditions disagree with the pair: $A_{18}[5] \ne A_{18}[10]$, $E_{17}[22] \ne E_{18}[22]$ and $W_{26}[8] = W_{26}[21]$ (printed for $W_{25}$). As in~\cite{GWZW26}, each enforced relation is checked as it holds in the pair, and all 93 hold on it. Timings ran on one performance core of an Apple M1 Pro, on which a scalar 39-step SHA-512 compression in portable C (expansion, 39 steps, feed-forward) takes 159.3\,ns; the machine was shared with four compute-bound processes during every timing, and all costs are ratios of times on it.

\paragraph{Row 17.} A dynamic program over the bits of $W_{17}$ counts $|S_{17}| = 1{,}404{,}469{,}248$ at the target of~\cite{GWZW26}, their count, and $2^{24.85}$ at the pair's own $W_{16}$. The bitwise check of row 17 agrees with a run of step 17 on both members on $2^{22}$ uniform words and on the 64 one-bit flips of each of 185 passing words.

\paragraph{The rule.} Each draw takes a random chaining value, $a_0$ and $a_1$, transplants $(a_2,a_3)$ and runs both members through 16 steps. The rule agrees with these runs on $2^{22}$ draws in each of four settings (uniform pairs, the fixed bits of $E_7$ right, admissible $a_3$, and admissible $a_3$ with the step-9 condition and $\mathcal{F}_{10}$), and with brute force on four classes, over all $2^{22}$ values of $a_2$ on 22 bits spread over the word. The price: the state part is exactly $2^{-21}$, and given the 20 fixed bits of $E_7$, $\mathcal{F}_{10}$ and $\mathcal{F}_{11}$ hold together with probability $2^{-13.0}$ (\cref{tab:price}). For a random admissible $a_3$ a class holds $2^{19}$ to $2^{32.5}$ entries; choosing $a_3$ with $A_4 \oplus a_3 = \lnot E_8$ on its free bits makes $a_2$ cancel in $h$ except through carries, and gives up to $2^{38.4}$. On $\eta = \hex{b8e7586be3fa23f5}$, the 2{,}275 admissible values of $a_3$ within three bits of one class's $E_7$ hold $2^{46.13}$ entries, an exact count over the classes scanned and a lower bound for the fibre.

\paragraph{The search.} With row 17 restricted to its conditions below bit 28, the search and a test of all 2.9 million entries of a class return the same sets on 32 solutions of row 16. With row 17 in full, the search returns a planted entry in each of 200 trials, and exactly the set that testing returns. The expected number of nodes, $\sum_k L(k)\,\pi(k-9)$ with $L(k)$ the live prefixes of length $k$ and $\pi$ the exact rate of row 17's conditions below a bit, is $2^{20.73}$ on the largest class; $2^{20.74}$ were measured. A survey of 640{,}000 pairs of class and fibre puts the floor of the cost per hit near $2^{13.8}$, with classes A and B among the best.

\paragraph{Hits per search.} A solution of row 16 is one sub-attempt $(\cv,a_0)$ at 1.213 compressions, which inverts $\Fodd$ (0.9993 solutions on average) and passes $\mathcal{F}_9$ with probability $2^{-6.000}$, so $C_{16} = 77.7$ compressions. The chaining value changes every 1{,}024 values of $a_0$, and the first block is not counted. Over 73{,}728, 36{,}864 and 256 searches, classes A, B and the largest gave $0.2145 \pm 0.0081$, $0.4385 \pm 0.0213$ and $1.61 \pm 0.89$ hits per search, against $0.2190$, $0.4525$ and $2.25$ expected. A rerun on 20{,}000 and 8{,}000 new solutions of row 16 gave $0.228$ and $0.453$.

\paragraph{The tail.} Steps 18 to 24 are measured by multilevel splitting, in eight replicates of 4{,}096 survivors per level with roots drawn uniformly from $S_{17}$, and give $82.97 \pm 0.03$ bits against 83.09 in~\cite[Table~15]{GWZW26}. Each of the 32{,}768 survivors of step 24 is then run through steps 25 to 38 with the pair's $(W_{10},W_{11})$, 64 entries of class A, 64 of class B, and 64 uniform pairs that meet the rule. Paired against the pair's words, the rate of steps 25 to 38 changes by $-0.007$, $-0.011$ and $-0.018$ bits, with 95\% bootstrap intervals within $[-0.08,+0.06]$. The tail from step 18 is $86.96 \pm 0.03$ bits with the pair's words, against $87.21 \pm 0.14$ in~\cite{GWZW26}; four further replicates with new seeds give $87.03 \pm 0.06$, and $86.99 \pm 0.03$ on entries of the list, a paired change of $+0.04$ bits on steps 25 to 38 (95\% interval $[-0.05,+0.13]$). Every survivor that passes step 38 collides. Like that of~\cite{GWZW26}, the tail is a splitting estimate whose standard errors neglect covariance across steps.

\section{Verification and Reproduction}\label[appendix]{app:verif}

\subsection{Evidence for the premises}

\paragraph{H1.} On 310{,}224 claws of the walk with two free words, $\mathcal{F}_9$ passes at $0.01617$ against $0.01587$, $\mathcal{F}_8$ at $0.07092$ against $0.07031$, both at $0.001160$ against $0.001116$, and the low 16 bits of $W_{17}$ are consistent with uniform ($\chi^2 = 65{,}830.2$ on $65{,}535$ degrees of freedom).

\paragraph{H5.} 37 hits were observed against 38.58 expected. Independence of distinct claw groups, within and across windows, is supported by the binomial behaviour of the $\mathcal{F}_8$ and $\mathcal{F}_9$ counts across windows ($\chi^2$ of 22.6 and 32.0 on 24 degrees of freedom).

\paragraph{H-cross-claw, pairs of hits.} The reference implementation declares a further premise, which the costs of this paper do not use. For hits $k$ and $l$ with different keys on different claws of one window, averaged over the attack's pairs of hits, the probability of a collision of $l$, given that a mate of $k$ follows the trail, is at most $2^{-8}$. In the 14 organizer-run replicas, 0 of 60{,}416 trials on another claw of a hit's chaining value kept the differences through row 19 (one-sided 99\% bound $7.62\cdot 10^{-5}$ per entry, $2^{-10.68}$ for eight mates per key), against 1 of 60{,}416 on random chaining values; 64 participant-run replicas give 2 against 5 through row 19 in 551{,}872 trials per arm. Row 19 is the deepest row measured. The bound $\Pr[U_k\cap U_l]\le\min(p_k,p_l)$ holds for any joint law of the tails. For keys on different claws of one window, the reference implementation replaces it with a measured premise, H-cross-claw (\cref{sec:listprem}): averaged over the attack's pairs of hits, a mate of $k$ that follows the trail makes a collision of $l$ no more likely than $2^{-8}$. Two hits of one window share only the chaining value and $W_{12},\dots,W_{17}$, and the premise is measured through row 19 only. It removes the $0.09$ bits of the bound and gives the reference implementation's figure of $2^{93.9}$ in the board's model; the costs below do not use it.

\paragraph{Selection of claws.} At sandbox scale, this correlation is measured for claws of the walk that pass $\mathcal{F}_9$ and entries in $\mathcal{F}_8$: on shared seeds and on windows screened as the attack screens them, their tail is at most $0.025$ bits lower than the estimator's at 95\%. Selection by the event $W_{17}=w_{17}$ is beyond that scale. Because $W_{17}$ shares $u = W_1$ with the row-16 equation, its uniformity over claws is the premise with the most direct route to a test; all 32 of its bits are consistent with uniform on 72{,}040 claws.

\subsection{Experiments of the filings}

The board runs each filing's declared experiments in a bounded, networkless sandbox and strips the participants' observations from what its judge sees, so an experiment establishes only what its exit status and the organizer's recomputation establish. The figures in the ledgers were measured by the participants outside the sandbox; the sandbox checks consistency and replicates the tail measurements.

\paragraph{The attack with four free words~\cite{PR1809}.} The reference implementation declares 16 experiments, all run by the organizer, with every reported value hard-asserted.
\begin{itemize}
\item \emph{Transplant and list.} The advice collides at 39 steps. For one entry of each of the 200 groups of equal $\kappa$, the rule of \cref{sec:list} holds, $W_9$ is the advice's value shifted by $\eta \boxminus \eta_0 = \hex{05cefde0}$, $W_{16}$ takes one value, and both members follow the advice trail through step 16. For random chaining values and random $(a_0,u)$ with a listed entry, the transplant reaches the pinned state, $W_1 = u$, and $W_{16} \boxminus g = \Fodd(u) \boxminus \Feven(a_0)$.
\item \emph{Small windows.} Four windows of $2^{21}$ evaluations are run in check mode, with the pass run on every claw that passes $\mathcal{F}_9$; they yield 233, 222, 256 and 232 distinct claws, of which 7, 11, 4 and 1 pass $\mathcal{F}_9$. Every such claw asserts its $W_9$ against a listed entry and $W_{16} = g$. On the first such claw in each window a planted key of each of the 200 groups is found at its counted cost. One window is run again under a work cap and stops exactly where the recorded work of its claws predicts; one window is run with the first-block screen, which discards 11 of 12 draws.
\item \emph{Key hits at a scaled event.} Over the entries of every 50th group and all claws of the four windows that pass $\mathcal{F}_9$, a hit is redefined to occur with probability $2^{-7}$ instead of $2^{-32}$, which gives 1{,}104 hits in 146{,}502 key events, against 1{,}144.5 expected.
\item \emph{Tail and mates.} Fourteen runs of the tail replica with one uniform listed entry per batch give 1{,}888 collisions, each recovered to a chaining value and a second block and verified by two compressions; these are SFS collisions, as in \cref{sec:tail}. Of 638 mates with $W_8 \in \mathcal{F}_8$ traced on the same claws, none follows the trail past row 19. None of the hits' values of $W_3$ lies in the sets where a mate could follow the trail, against 0.24 expected for a uniform $W_3$, and their per-bit $\chi^2$ is 24.7 on 32 degrees of freedom.
\item \emph{Pairs of hits and panels.} Each tail replica also runs 32 trials on another claw of each verified hit's chaining value and 32 on a random chaining value; the reference implementation predicts at most 17 trials through row 19 over the 14 runs and records 0, against 1 on random chaining values. On 12 real claws of the small windows, 3{,}714 trials are compared with virtual ones on the same chaining value at row 18: 1{,}830 against 1{,}873 passes. Two windows of $2^{22}$ evaluations on the organizer's seeds check the panel rates of distinct claws, of claws that pass $\mathcal{F}_9$ and of entries in $\mathcal{F}_8$ against bands calibrated on 2{,}400 fresh windows; these bands catch gross bias only.
\end{itemize}
The trial count, the yield of the work cap and the enumeration of rows 18 and 19 are run by the participants on predeclared public seeds and are not repeated in the sandbox. Independently of the filing, we rebuilt the list from its rule, counted its entries in $\mathcal{F}_8$ on 600 uniform claws, and checked the identities of \cref{lem:four} (\cref{sec:list,sec:listcost}). From the 2{,}160 per-window rows of the yield panel we recomputed the yield of the work cap and its bound. We also measured the tail of entries under $K_P$ and under the strict event of~\cite{GWZW26}, on 30 seeds per arm fixed before the runs (\cref{sec:gap}). We ran the attack up to step 17 on the saved claws of 108 windows: 83 keys meet $W_{17}=w_{17}$, against 78.3 expected at $2^{-32}$ per key, and \cref{app:tables} gives one of the resulting pairs from a real first block.

\paragraph{The attack with two free words~\cite{PR574}.} The published pair is pushed through the program's own claw path with the kernel built for $\cv_{\mathrm{pub}}$: the claw is returned, the three filters pass, and the pair collides. The identity of \cref{lem:transplant} and $W_{16} = g$ are checked on 255 random chaining values. Four windows cut to $2^{22}$ evaluations assert $F(a_0)=F(u)$, the parities and $W_{16}=g$ on every raw claw (3{,}265 distinct claws), with the pass counts of $\mathcal{F}_8$ and $\mathcal{F}_9$ and the uniformity of $W_{17}$ and $W_{18}$ modulo 256 hard-asserted. The active-bit enumeration of \cref{lem:side} is run and gates the result, the fibre sizes $|G_{16}|=2$ and $|G_{17}(g)|=1$ are enumerated, and 14 experiments replicate the tail on the organizer's public seeds, with every hit inverted to an SFS pair and verified by two compressions.

\subsection{Participant-run measurements}

Two of us ran four further measurements outside the sandbox. Every run is reported. In the tail measurements every hit was inverted to a chaining value, checked by two 39-step compressions, and its entry recovered from the colliding pair. The tail on the estimator's draws and the two counts over claws were pre-registered on 17 November, and the tail on real claws on 23 November, each before its counted runs.
\begin{itemize}
\item \emph{The tail on the estimator's draws.} The replica on 64 runs, under the law at which the tail is priced: $2^{-74.45}$, with the 95\% interval $[-74.49,\ -74.41]$ (\cref{sec:tail,tab:carrierrows}). Restricted to the parity class of the attack, $a_0$ even and $W_1$ odd, it gives $2^{-74.42}$ on 7{,}281 hits, with the interval $[-74.48,\ -74.37]$. The priced value $2^{-74.46}$ lies inside both intervals.
\item \emph{The tail on real claws.} The same replica with $W_8,\dots,W_{11}$ taken from real pairs of a claw and an entry. Each of 128 runs, with fresh identifiers, runs one window of the reference implementation on its attack path, with the first-block screen and the work cap, and takes the claws on which the window runs its pass, those that pass $\mathcal{F}_9$; an entry is drawn uniformly from the list and kept when $W_8\in\mathcal{F}_8$. These runs give 57{,}652 hits and $2^{-74.45}$. The replica's own law, run on the same identifiers, gives 58{,}025 hits and $2^{-74.44}$. A run's identifier seeds its sequential Monte Carlo, so each pair of runs shares the population of rows 18 to 23, and only $W_8,\dots,W_{11}$ differ. The mean per-run difference, real claws minus the replica's draws, is $-0.009$ bits, with the 95\% interval $[-0.025,\ +0.007]$ on 127 degrees of freedom. Every hit was checked a second time with the reference implementation's own compression, and its claw against the window's records. An earlier arm of 28 runs on windows without the first-block screen, whose rule for pricing was declared after 22 of its runs, gave $+0.03$ bits $[-0.02,\ +0.07]$; the two effects do not differ significantly ($-0.03$ bits, $[-0.08,\ +0.01]$). The selection of claws by the walk and by $\mathcal{F}_9$ does not lower the tail measurably. The estimator pins $W_{17}$, so selection by a hit on $W_{17}$ is not sampled.
\item \emph{The trial count on all claws.} Entries in $\mathcal{F}_8$ on all 72{,}040 claws that pass $\mathcal{F}_9$ in 80 full windows (\cref{sec:listcost}).
\item \emph{$W_{17}$ over claws.} On the same 72{,}040 claws, $W_{17}$ was computed by the program at a fixed entry and at an entry drawn per claw: the $\chi^2$ of the high 16 bits is $65{,}551.0$ and $65{,}267.2$ on $65{,}535$ degrees of freedom, that of the low 16 bits $65{,}329.1$ and $65{,}401.8$, and the per-bit $\chi^2$ is $31.77$ and $38.11$ on 32. The largest $|z|$ of a single bit, $2.54$, has probability about $0.30$ of occurring among 32 bits under uniformity. All 32 bits are consistent with uniform.
\end{itemize}
The pre-registrations, programs and per-run records are in the deposit of \cref{sec:repro}.

\subsection{Reproduction}\label{sec:repro}

Each filing cited here is the head commit of its pull request in~\cite{HashSmash}, which stays fetchable as \texttt{refs/pull/$N$/head} after the pull request is closed. The reference implementations are the following commits.
\begin{center}
\small
\begin{tabular}{@{}ll@{}}
Four free words~\cite{PR1809} & \texttt{e802a33d8435fd0d1130a6a60b1fe995cb098eb6} \\
Two free words~\cite{PR574} & \texttt{87dfdbd29c713d64d62aa84208ab29e013731002} \\
\end{tabular}
\end{center}
In each commit the package lies under \texttt{pareto/candidates/sha256/}: the claim, the argument, the experiment manifest and the programs it declares. The declared experiments are rerun with the experiment harness of the same commit, which fixes the container image and the organizer's seed protocol; the judge's review and the score are not part of this reproduction.

The measurements made for this report, with their programs and records, are deposited at~\href{https://doi.org/10.5281/zenodo.24556799}{10.5281/zenodo.24556799}: those of \cref{sec:gap,sec:pinning}, the price of free words on the published characteristics (\cref{app:price}), the list rebuilt from its rule with its counts over 600 uniform claws (\cref{sec:listcost}), the list and the search on 39-step SHA-512 (\cref{sec:sha512,app:sha512}), and the worked example of \cref{app:tables}. The deposit also holds the programs, pre-registrations and per-run records of the participant-run measurements, as the two archives of 17 November with their SHA-256 digests, and the per-window rows of the panels priced in~\cite{PR1809}.

\section{The Characteristic and the SFS Pairs}\label[appendix]{app:tables}

\Cref{tab:char,tab:pair} reproduce Tables~21 and~23 of~\cite{LZL+26}, in that paper's notation. In \cref{tab:char}, $x$ is the block $M'$ and $x'$ the block $M$ of \cref{tab:pair}; rows $-4$ to 6 and 27 to 38 consist of \texttt{=} only and are omitted, and rows 7 to 15 are the rows of the dense part (\cref{sec:transplant}) that carry conditions; rows 2 to 6 carry none. The additional conditions on rows 16 to 25 are given in Table~22 of~\cite{LZL+26}.

\begin{table}[!ht]
\centering
\caption{The differential characteristic for 39-step \sha{}, rows 7 to 26~\cite[Table~21]{LZL+26}.}\label{tab:char}
\setlength{\tabcolsep}{3pt}
\renewcommand{\arraystretch}{1.3}
\resizebox{\textwidth}{!}{%
\begin{tabular}{@{}c|c|c|c@{}}
\noalign{\hrule height 1pt}
$i$ & $\nabla A_i$ & $\nabla E_i$ & $\nabla W_i$ \\ \hline
7  & \texttt{================================} & \texttt{====1===0=100======0======0=====} & \texttt{================================} \\
8  & \texttt{=n==============================} & \texttt{=n0=0=001=011=11=0=100+=001=1=0=} & \texttt{=n==============================} \\
9  & \texttt{====n===n===u=n====n===u==u=====} & \texttt{=010u100n=unn==1000n11+001u=0111} & \texttt{====u===u==========n============} \\
10 & \texttt{================================} & \texttt{u0n1111n00u110110u10uuu1uuu0u0n1} & \texttt{=u==============================} \\
11 & \texttt{================================} & \texttt{1u1=u1110000n00011=1100n001=0=01} & \texttt{=====nn=======n===u===uu==uu=n==} \\
12 & \texttt{=====n==nn======================} & \texttt{1n111u00nu0010u00011100110010=0=} & \texttt{===========u=======uu===========} \\
13 & \texttt{==u=u==u=u==u==nn====u======n=uu} & \texttt{10u0u0n001uuuu01u01uu=nu0001u001} & \texttt{================================} \\
14 & \texttt{=n===========u============n=====} & \texttt{=001110=1100101=0==00=01====0===} & \texttt{================================} \\
15 & \texttt{===n=========uu========u==n=====} & \texttt{==0u0=1=1=0000==11=01100====0=11} & \texttt{================================} \\
16 & \texttt{=====u=n=======u================} & \texttt{===1====0=======1==001=n=======0} & \texttt{====u===u=1=0====1=n=0=====1=1==} \\
17 & \texttt{================================} & \texttt{n==1====n=====u=u==uuu=1======1n} & \texttt{=u==============================} \\
18 & \texttt{=u==============================} & \texttt{0======00=====100==000=0=======0} & \texttt{================================} \\
19 & \texttt{================================} & \texttt{0======01===u0111==011===1n====1} & \texttt{================================} \\
20 & \texttt{================================} & \texttt{=10===uu====1==u===011===11==0==} & \texttt{================================} \\
21 & \texttt{================================} & \texttt{=1====00====1==0==========1=====} & \texttt{================================} \\
22 & \texttt{================================} & \texttt{=u====10=======1===01===========} & \texttt{================================} \\
23 & \texttt{================================} & \texttt{=0==============================} & \texttt{================================} \\
24 & \texttt{================================} & \texttt{=1==============================} & \texttt{====1=nn=====1=n=1==============} \\
25 & \texttt{================================} & \texttt{================================} & \texttt{================================} \\
26 & \texttt{================================} & \texttt{================================} & \texttt{=n==============================} \\
\noalign{\hrule height 1pt}
\end{tabular}}
\end{table}

\begin{table}[!ht]
\centering
\caption{The SFS colliding message pair for 39 steps of \sha{}~\cite[Table~23]{LZL+26}; the words of $M'$ that differ from $M$ are $m_8$ to $m_{12}$.}\label{tab:pair}
\scriptsize
\setlength{\tabcolsep}{3pt}
\begin{tabular}{@{}l|llllllll@{}}
\toprule
$\cv$ & \texttt{b21e3e63} & \texttt{d12deef2} & \texttt{71a28861} & \texttt{1c40a0d4} & \texttt{15bb07fb} & \texttt{b326f54c} & \texttt{020b80dd} & \texttt{0e6446c1} \\
\midrule
\multirow{2}{*}{$M$} & \texttt{5ee50e16} & \texttt{b5a4f02b} & \texttt{a58b289e} & \texttt{f713de5d} & \texttt{3b96d2db} & \texttt{215da7e2} & \texttt{3674669a} & \texttt{2ca65d8d} \\
 & \texttt{e99c40e3} & \texttt{172e9a20} & \texttt{02e88746} & \texttt{07021c8f} & \texttt{e8c6e622} & \texttt{b06304c1} & \texttt{43660040} & \texttt{4891771f} \\
\midrule
\multirow{2}{*}{$M'$} & \texttt{5ee50e16} & \texttt{b5a4f02b} & \texttt{a58b289e} & \texttt{f713de5d} & \texttt{3b96d2db} & \texttt{215da7e2} & \texttt{3674669a} & \texttt{2ca65d8d} \\
 & \textcolor{red}{\texttt{a99c40e3}} & \textcolor{red}{\texttt{1fae8a20}} & \textcolor{red}{\texttt{42e88746}} & \textcolor{red}{\texttt{01003fbb}} & \textcolor{red}{\texttt{e8d6fe22}} & \texttt{b06304c1} & \texttt{43660040} & \texttt{4891771f} \\
\midrule
hash & \texttt{f42bbd82} & \texttt{0f057a39} & \texttt{267fcb51} & \texttt{e202bc25} & \texttt{81e4ed41} & \texttt{d74409dd} & \texttt{3247beb6} & \texttt{ec7ed559} \\
\bottomrule
\end{tabular}
\end{table}

\begin{table}[!ht]
\centering
\caption{The SFS colliding pair used as advice by the filings from 28 October 2026~\cite{PR1288}, on the characteristic of \cref{tab:char}; the words of $M'$ that differ from $M$ are $m_8$ to $m_{12}$, and both blocks give the listed output after 39 steps.}\label{tab:pair2}
\scriptsize
\setlength{\tabcolsep}{3pt}
\begin{tabular}{@{}l|llllllll@{}}
\toprule
$\cv$ & \texttt{aa79ff87} & \texttt{65c5bb49} & \texttt{d0a84bfd} & \texttt{90ccc398} & \texttt{fac64fb6} & \texttt{4ff7998a} & \texttt{b1af5ec5} & \texttt{aa1ed4d5} \\
\midrule
\multirow{2}{*}{$M$} & \texttt{8334d670} & \texttt{3813877e} & \texttt{f19d46d5} & \texttt{77556fc4} & \texttt{902f98c7} & \texttt{0500282c} & \texttt{a4235533} & \texttt{9100e365} \\
 & \texttt{e76f5fe2} & \texttt{5633326d} & \texttt{9d603088} & \texttt{06a64487} & \texttt{a94ae11e} & \texttt{b06304ad} & \texttt{43660038} & \texttt{d80d950f} \\
\midrule
\multirow{2}{*}{$M'$} & \texttt{8334d670} & \texttt{3813877e} & \texttt{f19d46d5} & \texttt{77556fc4} & \texttt{902f98c7} & \texttt{0500282c} & \texttt{a4235533} & \texttt{9100e365} \\
 & \textcolor{red}{\texttt{a76f5fe2}} & \textcolor{red}{\texttt{5eb3226d}} & \textcolor{red}{\texttt{dd603088}} & \textcolor{red}{\texttt{00a467b3}} & \textcolor{red}{\texttt{a95af91e}} & \texttt{b06304ad} & \texttt{43660038} & \texttt{d80d950f} \\
\midrule
hash & \texttt{8b102414} & \texttt{c3486344} & \texttt{a4e834ea} & \texttt{c86b8951} & \texttt{38947077} & \texttt{0d118794} & \texttt{f6eff9f4} & \texttt{50d39f69} \\
\bottomrule
\end{tabular}
\end{table}

\paragraph{A worked example from a real first block.} We ran the attack up to step 17 on the saved claws of the 80 screened windows of \cref{sec:listcost}, with the program of~\cite{PR1753} and the list regenerated by its own rule. The first block of each window is the program's draw from the window's seed, and its chaining value reproduces the claws on record: each of the 72{,}040 claws gives $W_{16}=g$ and $W_9\in\mathcal{F}_9$ with a listed entry. Testing all 3{,}462{,}264 keys on every claw finds 66 keys with $W_{17}=w_{17}$, against 58.1 expected at $2^{-32}$ per key. Their 198 entries all give $W_{16}=g$, $W_{17}=w_{17}$ and $W_9\in\mathcal{F}_9$, and 10 of them have $W_8\in\mathcal{F}_8$, against 13.9 expected at $9/128$. On the 25{,}126 claws of the 28 unscreened windows of the earlier arm on real claws (\cref{app:verif}), 17 keys meet $W_{17}=w_{17}$, against 20.3 expected, and 2 of their 45 entries have $W_8\in\mathcal{F}_8$.

\Cref{tab:worked} gives one of the ten. The first block is the 246th draw of 16 words from Python's \texttt{random.Random(802105)}, the first that passes the screen and the kernel caps of the program; its chaining value has a screen score of 3. The claw is the sixth of its window to pass $\mathcal{F}_9$. The state differences of the pair equal the advice's through step 17 and first differ at step 18, where the uncontrolled part begins, and the two 39-step compressions differ in 129 bits.

\begin{table}[h]
\centering
\caption{A worked example: window 802105 of the screened program, the claw $(a_0,u)$, and the listed entry $(a_2,a_3)$ on the fibre \hex{3cc9eebe}. $M_1'$ plays the role of $M'$ of \cref{tab:pair2} and $M_1$ that of $M$; the words of $M_1'$ that differ from $M_1$ are $m_8$ to $m_{12}$. The pair gives $W_{16}=g=\hex{bca1c321}$, $W_{17}=w_{17}=\hex{71e9ba98}$, $W_8=\hex{910f7c47}\in\mathcal{F}_8$ and $W_9=\hex{7891a1df}\in\mathcal{F}_9$. The last two rows are the 39-step compressions of $M_1$ and $M_1'$ from $\cv$; the tail did not hold.}\label{tab:worked}
\scriptsize
\setlength{\tabcolsep}{3pt}
\begin{tabular}{@{}l|llllllll@{}}
\toprule
\multirow{2}{*}{$M_0$} & \texttt{a3d3d8e6} & \texttt{575bc311} & \texttt{8618205d} & \texttt{87219b47} & \texttt{48309aa6} & \texttt{ff1c3e3b} & \texttt{25512f6d} & \texttt{251a1c0a} \\
 & \texttt{66b698c7} & \texttt{0c0a7b97} & \texttt{680588e6} & \texttt{453a17d7} & \texttt{a91b2200} & \texttt{78c58bbe} & \texttt{e84551ec} & \texttt{86cb43ad} \\
\midrule
$\cv$ & \texttt{f6584359} & \texttt{07a584f7} & \texttt{68269ddf} & \texttt{6b7f04d1} & \texttt{0205a465} & \texttt{a38daf62} & \texttt{5c21a5c8} & \texttt{d1a002b3} \\
\midrule
$a_0,u,a_1,a_2,a_3$ & \texttt{b1195c22} & \texttt{6fea9c33} & \texttt{d9c2aad9} & \texttt{e82695d4} & \texttt{ec1f3c56} & & & \\
\midrule
\multirow{2}{*}{$M_1$} & \texttt{b7d5d39f} & \texttt{6fea9c33} & \texttt{072c1ea7} & \texttt{966c22ab} & \texttt{1669efd0} & \texttt{eba6c968} & \texttt{65a49f9a} & \texttt{cdb74b8c} \\
 & \texttt{d10f7c47} & \texttt{7011b1df} & \texttt{00acefe1} & \texttt{3fa64445} & \texttt{a94ae11e} & \texttt{b06304ad} & \texttt{43660038} & \texttt{d80d950f} \\
\midrule
\multirow{2}{*}{$M_1'$} & \texttt{b7d5d39f} & \texttt{6fea9c33} & \texttt{072c1ea7} & \texttt{966c22ab} & \texttt{1669efd0} & \texttt{eba6c968} & \texttt{65a49f9a} & \texttt{cdb74b8c} \\
 & \textcolor{red}{\texttt{910f7c47}} & \textcolor{red}{\texttt{7891a1df}} & \textcolor{red}{\texttt{40acefe1}} & \textcolor{red}{\texttt{39a46771}} & \textcolor{red}{\texttt{a95af91e}} & \texttt{b06304ad} & \texttt{43660038} & \texttt{d80d950f} \\
\midrule
$M_1$ out & \texttt{6efb0e1d} & \texttt{955922fd} & \texttt{13348ba4} & \texttt{43794ef4} & \texttt{6c79bbfd} & \texttt{e2634478} & \texttt{fa581ac7} & \texttt{33b13785} \\
$M_1'$ out & \texttt{e8c7e668} & \texttt{d3d9ff60} & \texttt{8fb1e573} & \texttt{a4feb75a} & \texttt{6ad412ad} & \texttt{39004f1a} & \texttt{dd382142} & \texttt{1d360035} \\
\bottomrule
\end{tabular}
\end{table}

\section{Provenance of the Filings}\label[appendix]{app:provenance}

Every attack on \sha{} in this paper is a filing on the HashSmash board~\cite{HashSmash}, target \texttt{sha256-r39-prefix-v1}, exploratory lane; the measurements on 39-step SHA-512 of \cref{sec:sha512} are not a filing. Each filing is a credited derivative of an earlier one. \Cref{tab:lineage} lists the filings on the path that leads to the reference implementation (filing 466760ac~\cite{PR1809}), and \cref{tab:levers} summarizes the steps after 22 October. Scores in both tables are in the board's model. Times are the opening times of the pull requests (UTC); scores are the automated judge's, except that filings that were not valid show their claimed score, since the board scores only valid filings. Co-author fields are taken from the board's submission records. Filings and credits name participants by their board handles; the authors' handles are dariolina (Dariia Porechna), Jbenisek (Jeremy Benisek, also known as cyberaxe), mitchuski (Mitchell Travers), geeknik (Brian Carpenter), 0xplemonade (Pontus Olovsson), dkuthoore (Damian Kuthoore) and newjordan (Jordan Newman). sraiw (Sebastian R. Wain), who filed ed741369 and 793185f6~\cite{PR421}, and lamb356, a co-author of~\cite{PR574}, are not authors and are credited by handle. A filing is named by the first eight hex digits of its submission id, which is the id in the title of its pull request; the bibliography gives it in full.

\begin{table}[p]
\centering
\caption{The filings on the path to the reference implementation. Scores are $\log_2$ of work in the board's model, claimed scores of filings that were not valid in parentheses; a second handle is a co-author; times UTC, 2026. Uncited refilings and intermediate entries on the path are omitted.}\label{tab:lineage}
\setlength{\tabcolsep}{2pt}
\scriptsize
\begin{tabularx}{\textwidth}{@{}rll>{\raggedright}p{1.6cm}r>{\raggedright\arraybackslash}X@{}}
\toprule
PR & Id & Opened & Author & Score & Contribution \\
\midrule
193 & 6807f974 & 16 Oct 13:32 & dariolina & (121.67) & first filing of the row-16 claw: $A_0$, $A_1$ free at 39 steps \\
223 & ed741369 & 16 Oct 20:21 & sraiw & (121.07) & claw filing presented as a derivative of 6807f974 and f86da3c1 \\
306 & d9904432 & 17 Oct 05:38 & dariolina & 125.25 & the claw on our own dense base; first valid filing, dominated by generic search \\
421 & 793185f6 & 17 Oct 15:31 & sraiw & 120.46 & transplant of the published pair; 320-bit bijection; side conditions as active-bit theorems; dual-parity fibre and tail estimator \\
554 & 7bfdee03 & 19 Oct 00:06 & Jbenisek, dariolina & 118.6089 & bit-sliced 256-lane walk; exact executed ledger; 14 organizer-run tail replicas \\
574 & e536ea19 & 19 Oct 08:28 & lamb356, dariolina & 118.346 & shorter kernel (723 records); claw groups on 24 windows; advice construction checked in the sandbox \\
845 & 3dfc1284 & 20 Oct 23:04 & mitchuski & 118.276 & the program of 574 with a larger tail replica \\
894 & 0c307634 & 21 Oct 01:44 & geeknik & 118.146 & the same with a variance-balanced tail replica \\
1053 & d5732ab2 & 21 Oct 20:53 & 0xplemonade & 118.145 & the program of 894; window count from the direct bound on distinct claw groups \\
1126 & 64a8c927 & 22 Oct 16:41 & dkuthoore & 117.836 & cheaper charged bookkeeping of the walk (2.96 compressions per distinct claw against 3.65); tail $-75.47$ unchanged \\
\midrule
1209 & c8ea5149 & 27 Oct 06:56 & Jbenisek, dariolina & 116.895 & the walk of 64a8c927 on a re-signed advice pair \\
1288 & 29bb683c & 28 Oct 19:45 & Jbenisek, dariolina & 116.682 & advice re-solved on the published characteristic, in a good class of $\sig{0}(W_{11})$ \\
1494 & 69fd14e9 & 1 Nov 07:27 & 0xplemonade & 116.591 & recent-reset cut masks; 128-lane merge search; setup charged on rejected first blocks \\
1496 & 0f61eb7d & 1 Nov 19:59 & Jbenisek & 116.586 & the package of 69fd14e9 with three charged-operation rewrites of the walk \\
1504 & bd445130 & 4 Nov 09:28 & 0xplemonade & 116.556 & chaining-value screen before kernel generation \\
1620 & b1dc646b & 8 Nov 04:25 & Jbenisek & (104.589) & $a_2$ released over a list of 4,815 values; an experiment exceeded the organizer's time limit \\
1629 & 89ebfc13 & 8 Nov 07:00 & Jbenisek & 102.102 & $a_2$ and $a_3$ released on the fibre $\eta_0$; 115,010 entries; $W_{17}$ re-tried bit-sliced \\
1636 & 520d5200 & 8 Nov 22:13 & newjordan & 101.248 & rewrites of the bit-sliced pass \\
1665 & 062c2f28 & 9 Nov 15:37 & dariolina & 100.602 & $\mathcal{F}_8$ tested at $W_{17}$ hits only \\
1667 & 2d314ea2 & 9 Nov 16:44 & newjordan & 100.507 & the $W_{17}$ test written as $\Fodd(y)=r$ \\
1669 & 1d321afe & 9 Nov 17:24 & Jbenisek, dariolina & 99.890 & the 361,638-entry list \\
1671 & 41a413b1 & 9 Nov 18:20 & newjordan & 99.443 & $\Fodd(y)=r$ counted on the 361,638-entry list \\
1694 & 7adcc2ad & 11 Nov 03:38 & mitchuski & 98.989 & the pass on 18-bit prefixes \\
1696 & e42cd157 & 11 Nov 12:37 & mitchuski & 98.050 & prefix pass carried to the 821,520-entry list \\
1699 & 4ebf677a & 11 Nov 17:27 & Jbenisek, dariolina & 97.241 & the 821,520-entry list with the prefix pass; every hit counted \\
1708 & 1422ae31 & 12 Nov 10:15 & mitchuski & 94.443 & fibre re-based to \hex{3cc9eebe}; the list by rule, 9,701,592 entries \\
1712 & 1d92bc9c & 12 Nov 19:20 & Jbenisek & 94.225 & tail at the replica mean, declared transfer \\
1713 & 01733bbe & 12 Nov 21:08 & Jbenisek & 94.181 & the first-block screen on the new fibre \\
1725 & b2b62005 & 13 Nov 12:12 & mitchuski & 94.087 & entries as trials; bound on mates \\
1753 & a60f7594 & 17 Nov 02:16 & Jbenisek & 94.007 & asserted work cap; second-order terms re-priced \\
1764 & d5752476 & 17 Nov 15:49 & mitchuski & 93.917 & cross-claw pairs of hits, measured premise \\
1809 & 466760ac & 18 Nov 06:44 & Jbenisek & 93.898 & yield panel pooled to 2,160 windows; tail priced at the pooled mean; the reference implementation \\
\bottomrule
\end{tabularx}
\end{table}

\begin{table}[t]
\centering
\caption{From $2^{117.84}$ to $2^{93.90}$ in the board's model: the steps after 22 October 2026, with the first valid filing of each. Scores are the automated judge's ($\log_2$); the last two rows price pairs of hits on different claws under premise H-cross-claw, which the costs of \cref{sec:attack} do not use.}\label{tab:levers}
\setlength{\tabcolsep}{4pt}
\small
\begin{tabularx}{\textwidth}{@{}>{\raggedright\arraybackslash}Xllr@{}}
\toprule
Step & Filing & Date & Score \\
\midrule
Advice re-signed, then re-solved on the published characteristic, good class of $\sig{0}(W_{11})$ & \cite{PR1209,PR1288} & 27, 28 Oct & 116.682 \\
Walk bookkeeping; screen of the first block & \cite{PR1494,PR1496,PR1504} & 1, 4 Nov & 116.556 \\
$a_2$ and $a_3$ released on the fibre $\eta_0$, 115{,}010 entries, $W_{17}$ re-tried bit-sliced & \cite{PR1629} & 8 Nov & 102.102 \\
Rewrites of the pass & \cite{PR1636} & 8 Nov & 101.248 \\
$\mathcal{F}_8$ tested at $W_{17}$ hits only & \cite{PR1665} & 9 Nov & 100.602 \\
The test written as $\Fodd(y)=r$~\cite{PR1667}; 361{,}638 entries~\cite{PR1669} & \cite{PR1671} & 9 Nov & 99.443 \\
The pass on 18-bit prefixes & \cite{PR1694} & 11 Nov & 98.989 \\
821{,}520 entries under the prefix pass & \cite{PR1696,PR1699} & 11 Nov & 97.241 \\
Fibre re-based to \hex{3cc9eebe}, list by rule, 9{,}701{,}592 entries & \cite{PR1708} & 12 Nov & 94.443 \\
Tail priced at the replica mean~\cite{PR1712}; first-block screen on the new fibre & \cite{PR1713} & 12 Nov & 94.181 \\
Entries as trials; bound on mates & \cite{PR1725} & 13 Nov & 94.087 \\
Asserted work cap; panels and second-order terms re-priced & \cite{PR1753} & 17 Nov & 94.007 \\
Pairs of hits on different claws priced under a measured premise & \cite{PR1764} & 17 Nov & 93.917 \\
Yield panel pooled to 2{,}160 windows; tail priced at the pooled mean & \cite{PR1809} & 18 Nov & 93.898 \\
\bottomrule
\end{tabularx}
\end{table}

\paragraph{Credit lines.} A credit in this appendix names the participant whose filing first contains a step, as the notes of the filings record it. The filings were written with AI coding agents (see the paragraph on tools), so a credit does not divide a step between a participant and their agent, and it concerns only filings on the board. The filings~\cite{PR554,PR574} state, in identical words, that the attack is the construction of 793185f6 (\cite{PR421}) ``built on dariolina's row-16 claw idea \ldots\ and on her preregistered tail methodology'', and list dariolina as co-author. The first filing of the claw is 6807f974 (16 October, refuted); its note states the observation that $A_0$ and $A_1$ are both free at 39 steps and the claw on row 16, and ed741369 (16 October, refuted) presents itself as ``a credited derivative'' of it. The filing~\cite{PR421} attributes the two-block framework to~\cite{MNS13,LZL+26} and the distinguished-point method to~\cite{vOW99}. The characteristic and the SFS pair are the work of the authors of~\cite{LZL+26}. For the list of \cref{sec:list}, the notes of the filings agree on the following. The release of state words with $h$ fixed and the lists on the advice's fibre are Jbenisek's~\cite{PR1620,PR1629}. The pass without $\mathcal{F}_8$ is dariolina's~\cite{PR1665}. The form $\Fodd(y)=r$ of the $W_{17}$ test is newjordan's~\cite{PR1667}. The pass on 18-bit prefixes, carried to the 821{,}520-entry list in~\cite{PR1696}, the rule that generates the list and the re-basing of the fibre are due to Mitchell Travers (mitchuski)~\cite{PR1694,PR1708}, as are the count of entries as trials and the bound on mates~\cite{PR1725}. The recent-reset cut masks, the 128-lane merge search and the setup charge on rejected first blocks, all kept in~\cite{PR1809}, are 0xplemonade's~\cite{PR1494}, whose package~\cite{PR1496} ships and credits, as is the screen of the first block~\cite{PR1504}. The work cap is Jbenisek's~\cite{PR1753}. The premise on pairs of hits on different claws is Mitchell Travers's~\cite{PR1764}, and the final entry, which pools the yield panel and prices the tail at the pooled mean, is Jbenisek's~\cite{PR1809}. The exhaustive fibre counts $|G_{16}|=2$ and $|G_{17}(g)|=1$ are credited to both 73d1ef96~\cite{PR416} and 793185f6~\cite{PR421}.

\paragraph{Tools.} The packages were written with AI coding agents. By participant, their public notes declare: dariolina, Claude Opus 4.6 and 5, and Claude Fable 5.1, with Claude Code; sraiw and lamb356, Gemini 4 Argon with Antigravity; Jbenisek, Claude Opus 5.5 with Claude Code; mitchuski, Claude Opus 5.5 and Claude Fable 5.2 with the agentprivacy dual-agent harness, and Claude Code; geeknik, GLM-5.3 with ZCode; 0xplemonade, Claude Opus 5.5 with Claude Code, and \texttt{gpt-daybreak-blue-latest}, GPT-6 Astra and GPT 6.1 Sol with Codex; dkuthoore, Grok 4.7 with Grok Bot; newjordan, DeepSeek V4 Flash with Qoder CLI. A note declares the agent that produced its package; participants may have used other models in the research and preparation behind it, which the notes do not record.

\paragraph{Contributions to this report.} The authors are the participants whose filings lie on the path to the reference implementation, except sraiw and lamb356, who are thanked in the acknowledgements; each contributed the filings and steps credited above. Dariia Porechna drafted the report with AI assistance, ran the measurements that the report presents as its own (\cref{sec:listcost,sec:gap,sec:pinning}) and those on 39-step SHA-512 (\cref{sec:sha512}), which build on the list and the pass credited above, the tail on real claws of \cref{sec:verification} with Mitchell Travers's programs, and the worked example of \cref{app:tables}, and recomputed from the raw records every figure the report takes from the participant-run measurements. Mitchell Travers ran the other participant-run measurements of \cref{sec:verification} and the earlier arm on real claws, supplied their programs, pre-registrations and records, and supplied the saved claws on which the worked example was run. Jeremy Benisek supplied the per-window rows of the panels priced in~\cite{PR1753,PR1809} and checked the credit lines of his filings. Eigen Labs and Shielded Labs organize the board, run its judge and ran the organizer-run experiments of \cref{sec:verification}; they are not authors.

\end{document}